\documentclass[a4paper,twocolumn,10pt,accepted=2023-10-11,noamsfonts]{quantumarticle-ZH}
\pdfoutput = 1
\usepackage[numbers,sort&compress]{natbib}
\usepackage{bm}
\usepackage{bbm}

\usepackage{graphicx}
\usepackage{xcolor}
\usepackage[colorlinks=true,allcolors=blue]{hyperref}
\usepackage{url}
\usepackage{amssymb}
\usepackage{amsmath}
\usepackage{cleveref}
\usepackage{newtxtext}
\usepackage{mathtools}
\usepackage{newtxmath}

\DeclareMathOperator{\tr}{Tr}

\usepackage{pifont}

\usepackage[normalem]{ulem}
\begin{document}
\title{Completely Positive Map for Noisy Driven Quantum Systems Derived by Keldysh Expansion}
\author{Ziwen Huang}
\address{Superconducting Quantum Materials and Systems Center,
Fermi National Accelerator Laboratory (FNAL), Batavia, IL 60510, USA}
\email{zhuang@fnal.gov}
\author{Yunwei Lu}
\address{Department of Physics and Astronomy, Northwestern University, Evanston, IL 60208, USA}
\author{Anna Grassellino}
\address{Superconducting Quantum Materials and Systems Center,
Fermi National Accelerator Laboratory (FNAL), Batavia, IL 60510, USA}
\author{Alexander Romanenko}
\address{Superconducting Quantum Materials and Systems Center,
Fermi National Accelerator Laboratory (FNAL), Batavia, IL 60510, USA}
\author{Jens Koch}
\address{Department of Physics and Astronomy, Northwestern University, Evanston, IL 60208, USA}
\author{Shaojiang Zhu}
\address{Superconducting Quantum Materials and Systems Center,
Fermi National Accelerator Laboratory (FNAL), Batavia, IL 60510, USA}
\begin{abstract}
    Accurate modeling of decoherence errors in quantum processors is crucial for analyzing and improving gate fidelities. To increase the accuracy beyond that of the Lindblad dynamical map, several generalizations have been proposed, and the exploration of simpler and more systematic frameworks is still ongoing. In this paper, we introduce a decoherence model based on the Keldysh formalism. This formalism allows us to include non-periodic drives and correlated quantum noise in our model. In addition to its wide range of applications, our method is also numerically simple, and yields a CPTP map. These features allow us to integrate the Keldysh map with quantum-optimal-control techniques. We demonstrate that this strategy generates pulses that mitigate correlated quantum noise in qubit state-transfer and gate operations.
\end{abstract}
\maketitle
\section{Introduction}
\label{Sec:Intro}

The ubiquity of decoherence errors in current quantum computing platforms poses a bottleneck for performing error-correctable quantum computation \cite{Preskill_NISQ_review}. Further reducing these errors relies on the accurate modeling of them, which is challenging due to the presence of complicated drive and noise background. For example, if the quantum system is strongly driven, or the noise (quantum and classical) is correlated, the widely-used Lindblad master equation is generally not applicable \cite{Lidar_coarse_grain,Lidar_coarse_grain_drive,Groszkowski_Magnus_master,Dynamical_sweet_spot,Green_filer_func,Norris_filter_func,Didier_dynamical_sweet_spot}.

To obtain more accurate predictions,  recent research is exploring generalizations of the Lindblad master equation, where the constant damping operators and rates in the original form are replaced by time-dependent ones \cite{Lidar_coarse_grain,Lidar_coarse_grain_drive,Groszkowski_Magnus_master,Muller_Keldysh,Fogedby_field_theory_lindblad,TrushechkinUnified, Vcchini_lindblad_generalized,Rivas_coarse_graining,Alicki_coarse_graining,Brandes_coarse_graining}. Impressively, after a more careful treatment of the bath degrees of freedom than that in the Lindblad formalism, the master equation not only becomes compatible with drives and correlated noise, but also maintain the property of generating completely positive and trace-preserving (CPTP) maps \cite{Lidar_coarse_grain,Lidar_coarse_grain_drive,Vcchini_lindblad_generalized,Rivas_coarse_graining,Alicki_coarse_graining,Brandes_coarse_graining}. In addition to this route, a formalism based on filter functions has been developed to model errors caused by correlated noise. This formalism can predict the sensitivity of the driven system to noise at different frequencies \cite{Green_filer_func,Green_filter_func_high_order,Bluhm_filter_functions,Bluhm_filter_functions_PRR,Oliver_flux_qubit_dd,Norris_filter_func,Viola_noise_spectroscopy,Zeng_filter_func}. As a comparison, the filter-function method usually requires fewer integrals, and has a clearer physical picture of how noise at difference frequencies contributes differently. However, this method mostly focuses on classical (or dephasing) noise, and does not always guarantee the CPTP character of the map.

In this paper, we present a decoherence model which combines the advantages of the two routes mentioned above, and is tailored for optimizing gate operations. Our method belongs to the filter-function category, while the Keldysh technique \cite{Muller_Keldysh,Shnirman_Keldysh_dephasing} used here extends the scope of the formalism in Ref.~\cite{Green_filer_func,Bluhm_filter_functions} to quantum noise. Furthermore, the map derived by this method is guaranteed to be CPTP, after a special secular approximation to the filter functions. Such approximation also significantly simplifies the calculation, which further allows us to explore error-mitigation strategies by integrating our method with the technique of quantum optimal control \cite{Leung_optimal_control,Norris_filter_func}. Using a few examples, we show that such a combination can generate pulses that suppress decoherence errors induced by correlated quantum noise.

The paper is structured as follows. In Sec.\ \ref{Sec:DerKeldysh}, we outline the derivation of the Keldysh map. The main results are summarized in Eqs.~\eqref{map} and \eqref{CPTP}. In Sec.\ \ref{Sec:Appl}, we apply our method to a variety of quantum systems, which not only reproduces some familiar results, but also extends the prediction of decoherence errors to several less familiar situations.  In Sec.\ \ref{Sec:optimal}, we integrate the Keldysh method with the quantum-optimal-control technique, and demonstrate improvement of gate and state-transfer fidelities via the optimization of drive pulses.

\section{Deriving Keldysh Maps}
\label{Sec:DerKeldysh}

\subsection{Formal Keldysh expansion}
\label{Subsec:formal}
We start by deriving the formal map for the qubit density matrix using the Keldysh expansion. The Hamiltonian of the full system is
\begin{align}
    \hat{H}(t) = \hat{H}_s(t) +  \hat{H}_B + \epsilon\hat{H}_{I},
    \label{eq:Hamiltonian}
\end{align}
where $\hat{H}_s(t),\hat{H}_B,\hat{H}_I$ denote the Hamiltonians for a driven quantum system, bath, and interaction. The system Hamiltonian $\hat{H}_s(t) = \hat{H}_{s0} + \hat{H}_d(t)$ comprises the static Hamiltonian $ \hat{H}_{s0}$ and the drive operator $\hat{H}_d(t)$. For simplicity, we specify the system-bath interaction by $\hat{H}_{I} = \hat{x}\hat{\eta}$, where $\hat{x}$ and $\hat{\eta}$ are the system and bath operators, respectively\footnote{In the main text, we focus on time-independent $\hat{x}$ and $\hat{\eta}$ for a concise description, but the derivation in Sec.\ \ref{Sec:DerKeldysh} can be straightforwardly generalized to cases where the system operator is time-dependent. Such a situation becomes important in several experiments, e.g., Refs.~\cite{Dynamical_sweet_spot_exp,Didier_dynamical_sweet_spot_exp}}. The small dimensionless parameter $\epsilon$ is used to keep track of the order.

In Eq.~\eqref{eq:Hamiltonian}, we assume that the interaction is weak and can be treated perturbatively. To conveniently perform the perturbative calculation, we first move to the interaction picture with the unperturbed propagator $\hat{U}_0(t) = \hat{U}_s(t)\otimes \hat{U}_B(t)$, where the partial propagator for the system is $\hat{U}_s(t) = \mathcal{T}\exp[-i\int_0^{t} dt' \hat{H}_s(t')]$ and bath $\hat{U}_{B}(t) = \exp[-i\hat{H}_Bt]$. In this rotating frame, the reduced qubit density matrix at time $\tau$ is:
\begin{align}
    \tilde{\rho}_s(\tau) = \mathrm{Tr}_B\big\{ \tilde{U}_I(\tau)\, \tilde{\rho}_s (0)\!\otimes\! \tilde{\rho}_B(0)\,  \tilde{U}^\dagger_I(\tau) \big\}.
    \label{eq:rhoqt}
\end{align}
Here, the interaction-picture propagator is given by $\tilde{U}_{I}(\tau)=\mathcal{T}\exp[-i\int_0^\tau dt\epsilon\tilde{H}_I(t)]$ and $\tilde{H}_I(t) = \hat{U}^\dagger_0(t)\hat{H}_I\hat{U}_0(t)$ is the system-bath coupling term in the interaction picture. We use $\tilde{\rho}_s(0)$ and $\tilde{\rho}_B(0)$ to denote the initial partial density matrices for the system and bath. Note that in this work, we assume that there is no entanglement between the system and bath initially, and the bath is prepared in its thermal equilibrium $\hat{\rho}_{B,\mathrm{eq}}$. 

To evaluate this formal expression, we expand $\tilde{U}_I(\tau) = \sum_\nu \tilde{U}_I^{(\nu)}(\tau)$ as a Dyson series, where the $\nu$th term $\tilde{U}_I^{(\nu)}(\tau)$ is given by
\begin{align}
    \tilde{U}^{(\nu)}_I(\tau)\!=&(-i)^{\nu}\epsilon^\nu\!\int_0^\tau\! dt_1 \tilde{H}_I(t_1)\int_0^{t_1} dt_2\tilde{H}_I(t_2) \nonumber\\
    &\cdots\times\int_0^{t_{\nu-1}} dt_\nu\tilde{H}_I(t_\nu).
    \label{eq:prop_nu}
\end{align}
Inserting this into Eq.~\eqref{eq:rhoqt}, we further expand the qubit density matrix as
\begin{align}
    \tilde{\rho}_s(\tau) = \sum_{\nu',\nu''}\mathrm{Tr}_B\big\{ \tilde{U}^{(\nu')}_I(\tau)\, \tilde{\rho}_s (0)\!\otimes\! \tilde{\rho}_B(0)\,  \tilde{U}^{(\nu'')\dagger}_I(\tau) \big\}.
    \label{eq:rhoqt2}
\end{align}
To simplify this expression, we define the $\nu$th-order map and the sum map as
\begin{align}
    &\mathbf{\Pi}^{(\nu)}(\tau)\,\boldsymbol{\cdot} \equiv \sum_{\nu'+\nu''=\nu}\mathrm{Tr}_B\big\{ \tilde{U}^{(\nu')}_I(\tau) [\boldsymbol{\cdot}  \otimes \tilde{\rho}_B(0)]\,  \tilde{U}^{(\nu'')\dagger}_I(\tau) \big\}, \nonumber\\
    &\mathbf{\Pi}(\tau) = \sum_{\nu\in\mathbb{N}}\mathbf{\Pi}^{(\nu)}(\tau) \label{eq:Pinut}
\end{align}
which casts Eq.~\eqref{eq:rhoqt2} into
\begin{align}
    \tilde{\rho}_s(\tau) 
    =\mathbf{\Pi}(\tau)\tilde{\rho}_s(0).\label{eq:Pidef}
\end{align}
Above, $\mathbf{\Pi}^{(\nu)}(\tau)$ only contains terms of order $\epsilon^\nu$. For $\nu=0$, we have $\mathbf{\Pi}^{(0)}(t) = \mathbf{I}_s$ (i.e., the superoperator-identity acting on density matrices  $\mathbf{I}_s\tilde{\rho}_s = \tilde{\rho}_s$), while higher-order terms describe the decoherence effects due to the system-bath coupling.

Although we can in principle use  Eq.~\eqref{eq:Pinut} to calculate the map $\mathbf{\Pi}(\tau)$ to arbitrary order, it is usually not the most convenient quantity to extract physical measurables from, according to the discussion in Refs.~\cite{Muller_Keldysh,Clerk_non_Gaussian,Huang_Keldysh}.  Instead, we follow the Keldysh theory and define the self-energy 
\begin{align}
    \mathbf{\Sigma}(\tau) \equiv \ln[\mathbf{\Pi}(\tau)],\label{SigmaPi}
\end{align}
where redundant terms in higher-order expansions can be conveniently identified, and the derivation of quantities such as relaxation rates is easier \cite{Muller_Keldysh,Clerk_non_Gaussian,Huang_Keldysh}. Below, we will focus on $\mathbf{\Sigma}(\tau)$. Similar to $\mathbf{\Pi}(\tau)$, the self-energy $\mathbf{\Sigma}(\tau)$ can be expanded in powers of $\epsilon$ by $\mathbf{\Sigma}(\tau) = \sum_{\nu}\mathbf{\Sigma}^{(\nu)}(\tau)$, where $\mathbf{\Sigma}^{(\nu)}(\tau)$ can be derived from $\mathbf{\Pi}^{(\nu)}(\tau)$ by a Taylor expansion. For example, the lowest two orders are related by
\begin{align}
    \mathbf{\Sigma}^{(1)}(\tau) = \mathbf{\Pi}^{(1)}(\tau),\quad \mathbf{\Sigma}^{(2)}(\tau) = \mathbf{\Pi}^{(2)}(\tau) - \frac{1}{2}\Big[\mathbf{\Pi}^{(1)}(\tau)\Big]^2.\nonumber
\end{align}
These relations can be further simplified, if we assume that the noise has a zero mean, $\mathrm{Tr}_B\{\tilde{\eta}(t)\tilde{\rho}_B(0)\}=0$. In that case, the first-order map $\mathbf{\Pi}^{(1)}(\tau)$ vanishes, resulting in the following simplified relations
\begin{align}
    \mathbf{\Sigma}^{(1)}(\tau)=0,\quad \mathbf{\Sigma}^{(2)}(\tau)=\mathbf{\Pi}^{(2)}(\tau).\label{pisigma2}
\end{align}

\subsection{Second-order truncation}
\label{Subsec:second-order}

For most experiments involving gate operations and state transfer, it is usually sufficient to estimate the decoherence error up to leading order. In the following, we focus on the leading-order self-energy $\mathbf{\Sigma}^{(2)}(\tau)$. Explicitly, the second-order self-energy takes the form
\begin{align}
    \mathbf{\Sigma}^{(2)}(\tau)\tilde{\rho}_s(0) = &\,\mathrm{Tr}_B\Big\{\tilde{U}^{(2)}_I(\tau)\tilde{\rho}_s (0)\!\otimes\! \tilde{\rho}_B(0)\nonumber\\
    &\qquad+ \tilde{\rho}_s (0)\!\otimes\! \tilde{\rho}_B(0)\tilde{U}^{(2)\dagger}_I(\tau)\nonumber\\
    &\qquad+ \tilde{U}^{(1)}_I(\tau)\tilde{\rho}_s (0)\!\otimes\! \tilde{\rho}_B(0)\tilde{U}^{(1)\dagger}_I(\tau)\Big\}.
    \label{Pi2}
\end{align}
With the knowledge of the noise spectrum, we can further expand the right-hand side of Eq.~\eqref{Pi2}. For example, the first term can be expressed as
\begin{align}
    &\mathrm{Tr}_B\Big\{\tilde{U}^{(2)}_I(\tau)\tilde{\rho}_s(0)\!\otimes\! \tilde{\rho}_B(0)\Big\}\nonumber\\
    =&\,(-i)^2\!\int_0^\tau\!\!dt_1\!\int_0^{t_1}\!\!dt_2\,\tilde{x}(t_1)\tilde{x}(t_2)\tilde{\rho}_s(0) \nonumber\\
    &\qquad\quad\times\epsilon^2\mathrm{Tr}_B\{\tilde{\eta}(t_1)\tilde{\eta}(t_2)\hat{\rho}_{B,\mathrm{eq}}\},\nonumber\\
    =&\,-\int_{-\infty}^{\infty}\frac{d\omega}{2\pi}S_B(\omega)\int_0^\tau\!\!dt_1\!\int_0^{t_1}\!\!dt_2\nonumber\\
    &\quad\times \tilde{x}(t_1)\tilde{x}(t_2)\tilde{\rho}_s(0) e^{-i\omega (t_1-t_2)},\label{eq:Pioneterm}
\end{align}
and the other two can be derived similarly. These integrals can be conveniently summarized by four Keldysh diagrams, which we show and explain in Appendix \ref{ap:filter}. In the equation above, $S_B(\omega)\equiv \epsilon^2\int_{-\infty}^{\infty}dt \mathrm{Tr}_B\{\hat{\rho}_{B,\mathrm{eq}}\Tilde{\eta}(t)\tilde{\eta}(0)\}\exp(i\omega t)$ is the noise spectrum, and the interaction-picture operators are derived as $\tilde{x}(t) = \hat{U}^\dagger_s(t)\hat{x}\hat{U}_s(t)$ and $\tilde{\eta}(t) = \hat{U}^\dagger_B(t)\hat{\eta}\hat{U}_B(t)$. 

Given the information of the noise spectrum $S_B(\omega)$ and system propagator $\hat{U}_s(t)$, we can use Eqs.~\eqref{Pi2} and \eqref{eq:Pioneterm} to calculate the approximated dynamical map \cite{Lidar_coarse_grain,Alicki_coarse_graining,Brandes_coarse_graining,Rivas_coarse_graining}, i.e.\footnote{For a noise bath initiated as $\rho_{B,\mathrm{eq}}=\sum_\mu \lambda_\mu\vert \mu\rangle\langle \mu\vert$ ($\lambda_\mu>0$), the formal map \eqref{fullwave} without any approximation or truncation is already formally CPTP. (The proof of this is analogous to that in \cite{Lidar_coarse_grain} for a static system.) Note that sufficiently accurate calculation of the integrals \eqref{map_full} may be required toward a numerical CPTP map.},
\begin{align}
    \mathbf{\Pi}(\tau)\approx \exp\Big[\mathbf{\Sigma}^{(2)}(\tau) \Big].\label{fullwave}
\end{align}
This calculation is reminiscent of the filter-function method shown in Refs.~\cite{Green_filer_func,Green_filter_func_high_order,Norris_filter_func,Bluhm_filter_functions,Bluhm_filter_functions_PRR}. For example, the double integral  
$$\int_0^t\int_0^{t_1}dt_1dt_2\tilde{x}(t_1)\tilde{x}(t_2)\exp[-i\omega(t_1-t_2)]$$ is closely related to the filter functions studied there. (See Appendix \ref{ap:classicalconnection} for a detailed discussion of the connection to these theories.) For comparison, the time ordering of the two coupling operators $\tilde{\eta}(t_1)$ and $\tilde{\eta}(t_2)$ in the Keldysh expansion resolves the asymmetric noise spectrum for a non-classical noise source. 

However, evaluating the triple integral in Eq.~\eqref{eq:Pioneterm} accurately is usually not numerically efficient. In the next section, we show that appropriate approximations can simplify the calculation.

\subsection{Fourier expansion and secular approximation}
\label{Subsec:Fourierexp}
Our strategy for solving the aforementioned problem is  based on the Fourier expansion of 
\begin{align}
    \tilde{x}(t) = \sum_k \tilde{x}_{k}\exp(-ik\omega_p t),\label{eq:Fourier}
\end{align}
where we define the fundamental frequency $\omega_p = 2\pi/\tau$. Inserting it into Eqs.~\eqref{eq:Pioneterm} and \eqref{Pi2}, we find
\begin{align}
    \mathbf{\Sigma}^{(2)}(\tau)\tilde{\rho}_s(0) 
    =&\,-\sum_{kk'}\tilde{x}_k\tilde{x}_{k'}\tilde{\rho}_s(0)\!\!\int_{-\infty}^{\infty}\!\!\frac{d\omega}{2\pi}S_B(\omega)I_{-k,k'}(\omega)\nonumber\\
    &-\!\sum_{kk'}\tilde{\rho}_s(0)\tilde{x}_{k'}\tilde{x}_{k}\!\!\int_{-\infty}^{\infty}\!\!\frac{d\omega}{2\pi}S_B(\omega)I^*_{k,-k'}(\omega)\nonumber\\
    &+\!\sum_{kk'}\tilde{x}_{k}\tilde{\rho}_s(0)\tilde{x}_{k'}\label{map_full}\\
    &\times\int_{-\infty}^{\infty}\!\!\frac{d\omega}{2\pi}S_B(\omega)[I^*_{k,-k'}(\omega)+I_{-k',k}(\omega)].\nonumber
\end{align}
Here, $I_{k,k'}(\omega)$ is the filter function defined by
\begin{align}
        I_{k,k'}(\omega) \equiv \int_0^\tau\!\!dt_1\!\int_0^{t_1}\!\!dt_2 \,e^{i(k\omega_p-\omega)t_1-i(k'\omega_p-\omega)t_2}.
        \label{eq:ikk'}
\end{align}
(The analytical evaluation of this integral is discussed in Appendix \ref{ap:filter}.) For $k=k'$ and $k\neq k'$,  the filter functions $I_{k,k'}(\omega)$ behave differently, which is worth careful inspection.

\begin{figure}
    \centering
\includegraphics[width=8cm]{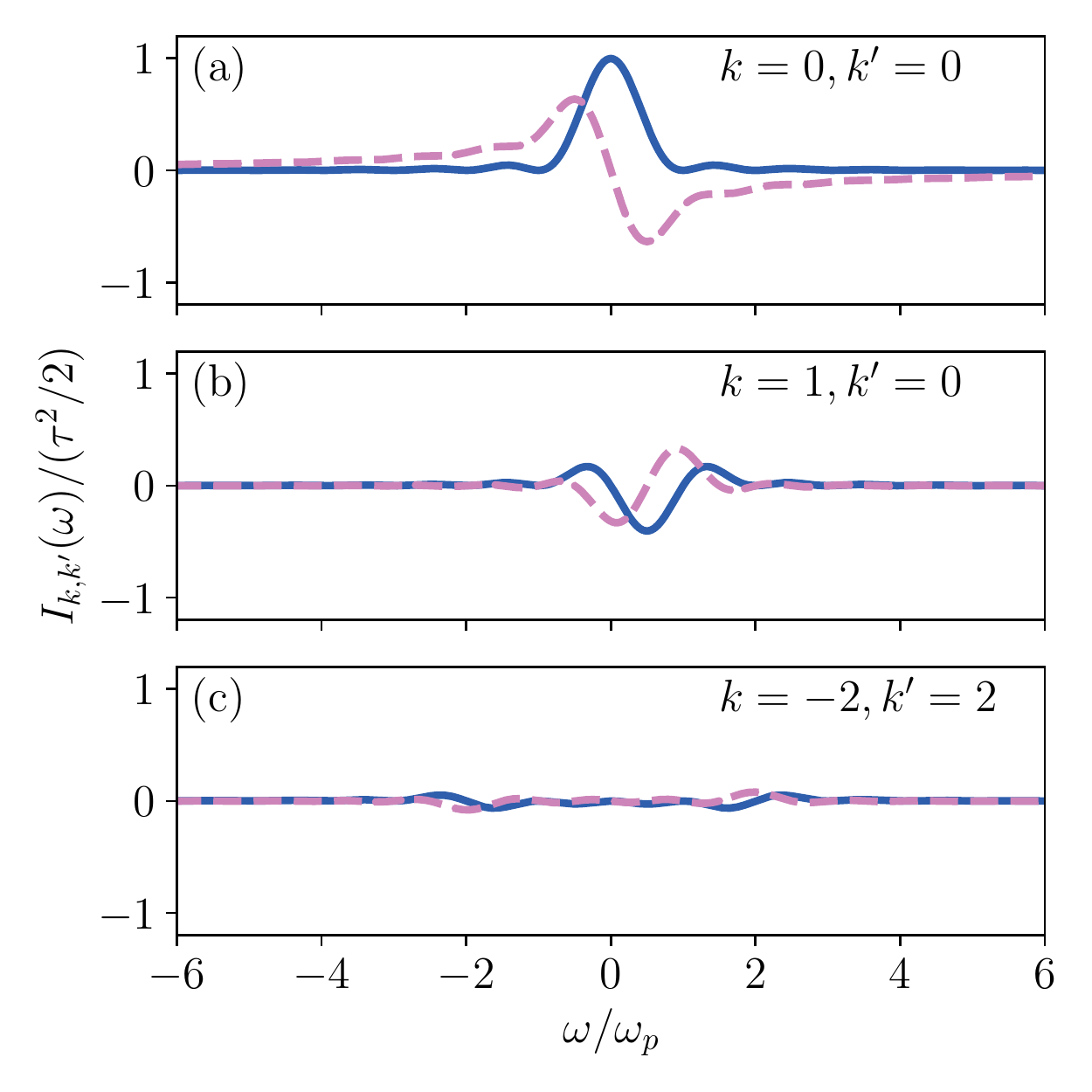}
    \caption{The filter functions $I_{k,k'}(\omega)$. The real and imaginary parts are shown as solid and dashed curves, respectively. From (a) to (c), $|k-k'|$ is chosen as  0, 1, and 4, respectively.}
    \label{fig:Filter_Func}
\end{figure}

\textit{Diagonal filter functions.--} For $k=k'$, the filter functions $I_{k,k'}(\omega)$ can be cast into the following form
\begin{align}
    I_{k,k}(\omega)= K^R(\omega-k\omega_p) + iK^I(\omega-k\omega_p), \label{IkkKRKI}
\end{align}
where the real and imaginary parts are given by:
\begin{align}
        K^R(\omega) = \frac{\tau^2}{2}\mathrm{sinc}^2\left(\frac{\omega \tau}{2}\right),\,\,\, K^I(\omega) = -\frac{\tau}{\omega}[1-\mathrm{sinc}(\omega\tau)].\label{eq:KRKI}
\end{align}
We illustrate the behavior of $K^R(\omega)$ (solid line) and $K^I(\omega)$ (dashed line) in Fig.~\ref{fig:Filter_Func}. Panel (a) shows the real and imaginary parts of  $I_{0,0}(\omega)$, respectively. Visibly, the function $K^R(\omega)$ has a predominant peak located at $\omega = 0$. According to Eq.~\eqref{eq:KRKI}, the width of this peak is $\sim 2\pi/\tau$. Different from the real part, $\mathrm{Im}\{I_{0,0}(\omega)\} = K^I(u)$ flips its sign at $\omega=0$, showing both a peak and a valley. Compared to $K^R(\omega)\lesssim 2|\omega|^{-2}$ in the limit $|\omega|\gg \omega_p$, $K^I(\omega)$ decays more slowly as $|K^I(\omega)|\sim \tau|\omega|^{-1}$.

\textit{Off-diagonal filter functions.--} 
The off-diagonal elements $I_{k,k'}(\omega)$ ($k\neq k'$) have three distinctive behaviors: (1) their amplitudes are smaller, and decrease as $|I_{k,k'}(\omega)|\lesssim\tau^2/(2\pi|k-k'|)$ for larger $|k-k'|$ {(see Appendix \ref{ap:filter})}. In the limit of $|\omega-k\omega_p|,|\omega-k'\omega_p|\gg |k-k'|\omega_p$, they have a fast  $|\omega-k\omega_p|^{-2}$ decay. (2) The peaks (valleys) are spread over a wider frequency range; the width of this frequency range is approximately $|k-k'|\omega_p$. (3) The off-diagonal filter functions elements have net-zero integrals, namely \begin{align}
    \int_{-\infty}^{\infty}\!\! \frac{d\omega}{2\pi} I_{k,k'}(\omega) = &\,\int_0^\tau\!\!dt_1\!\int_0^{t_1}\!\!dt_2 \,e^{ik\omega_pt_1-ik'\omega_pt_2}\delta(t_1-t_2)\nonumber\\
    =&\,\frac{1}{2}\int_0^\tau dt_1 e^{-i(-k+k')\omega_pt_1} = \frac{1}{2}\tau\delta_{k,k'}.\!\label{eq:area}
\end{align}
All these properties can be observed in Fig.~\ref{fig:Filter_Func} (a)-(c) for different values of $|k-k'|$.

Based on these three features, we arrive at the following conclusion: if variations of $S_B(\omega)$ are insignificant over the frequency scale of a few $\omega_p$, the off-diagonal elements of 
\begin{align}
    \phi_{k,k'}\equiv \int_{-\infty}^{\infty} \frac{d\omega}{2\pi} I_{k,k'}(\omega)S_B(\omega)
\end{align}
with $k-k'\neq 0$  have negligible amplitude. We justify this claim in three steps. First, for large $|k-k'|$, the amplitudes of $I_{k,k'}(\omega)$ are small, rendering a negligible $\phi_{k,k'}$. Second, for terms with small but nonzero $|k-k'|$, the slow-varying $S_B(\omega)$ allows us to treat it as quasi-constant. Third, using the property of net-zero area in Eq.~\eqref{eq:area},  the integral $\phi_{k,k'}$ vanishes for small but non-zero $|k-k'|$. Since $\phi_{k,k'}$ are the coefficients of terms shown in Eq.~\eqref{map_full}, we conclude that all off-diagonal terms are less important than the diagonal ones in that expansion, if the spectrum is sufficiently smooth at the resolution determined by $\omega_p$.

After neglecting the terms with off-diagonal filter functions, we simplify Eq.~\eqref{map_full} to
\begin{widetext}
\begin{align}
        \mathbf{\Sigma}^{(2)}(\tau)\tilde{\rho}_s(0)\approx \mathbf{\Sigma}^{(2)}_{\mathrm{CP}}(\tau)\tilde{\rho}_s(0) 
=&\,\sum_{k\in\mathbb{Z}}\Bigg[\tilde{x}_{k}\tilde{\rho}_s(0)\tilde{x}^\dagger_{k}-\frac{1}{2}\tilde{x}^\dagger_{k}\tilde{x}_{k}\tilde{\rho}_s(0)-\frac{1}{2}\tilde{\rho}_s(0)\tilde{x}^\dagger_{k}\tilde{x}_{k}\Bigg]\Big[\int_{-\infty}^{\infty}\!\!\frac{d\omega}{2\pi}S_B(\omega)2K^R(\omega-k\omega_p)\Big]\nonumber\\
    &-i\!\sum_{k\in\mathbb{Z}}\Big[ \tilde{x}^\dagger_k\tilde{x}_k\tilde{\rho}_s(0) - \tilde{\rho}_s(0)\tilde{x}^\dagger_k\tilde{x}_k\Big]\Big[\int_{-\infty}^{\infty}\!\!\frac{d\omega}{2\pi}S_B(\omega)K^I(\omega-k\omega_p)\Big]\label{map}
\end{align}
\end{widetext}
where 
$\mathbf{\Sigma}^{(2)}_{\mathrm{CP}}(\tau)$ is the simplified second-order self-energy. This step resembles the secular approximation performed in the derivation of the Lindblad master equation -- in both cases, small off-diagonal terms are neglected. Since the coefficient $\int_{-\infty}^{\infty} (d\omega/2\pi)S_B(\omega)2K^R(\omega-k\omega_p)$ is strictly positive, the self-energy $\mathbf{\Sigma}_{\mathrm{CP}}^{(2)}(\tau)$ has the form of a Lindbladian (up to an extra time dimension) \cite{Lindblad}. Then, according to Ref.~\cite{Lindblad}, the exponential of $\mathbf{\Sigma}^{(2)}_{\mathrm{CP}}(\tau)$ yields a CPTP map
\begin{align}
    \mathbf{\Pi}(\tau)\approx\exp\Big[\mathbf{\Sigma}_{\mathrm{CP}}^{(2)}(\tau)\Big].\label{CPTP}
\end{align}
In the following, we refer to $\Tilde{x}_k= [\int_0^\tau dt \Tilde{x}(t)\exp(ik\omega_pt)]/\tau$ and  $\omega_k = k\omega_p$  as the \textit{filter operator} and its corresponding \textit{filter frequency}, respectively.

According to the justification of the secular approximation, the map \eqref{CPTP} tends to be more accurate for smoother noise spectrum spectra. However, we observe that even for an $S_B(\omega)$ that exhibits strong peaks, the magnitude of the terms in Eq.~\eqref{map_full} with diagonal filter functions can still dominate those with the off-diagonal ones. As a result, the secular map \eqref{CPTP} is found to qualitatively agree with the full map for many common noise spectra, including those showing strong peaks. This is illustrated in Sec.\ \ref{Subsec:1/f} for the example of $1/f$ noise. Therefore, although we find it challenging to quantify the magnitude of the approximation error for arbitrary noise spectra, we still adopt Eq.~\eqref{CPTP} for an estimation if an extreme spectrum is considered; then, the agreement between the secular and full maps can be checked for validation.

\begin{figure}[h!]
    \centering
    \includegraphics[width = 7.5cm]{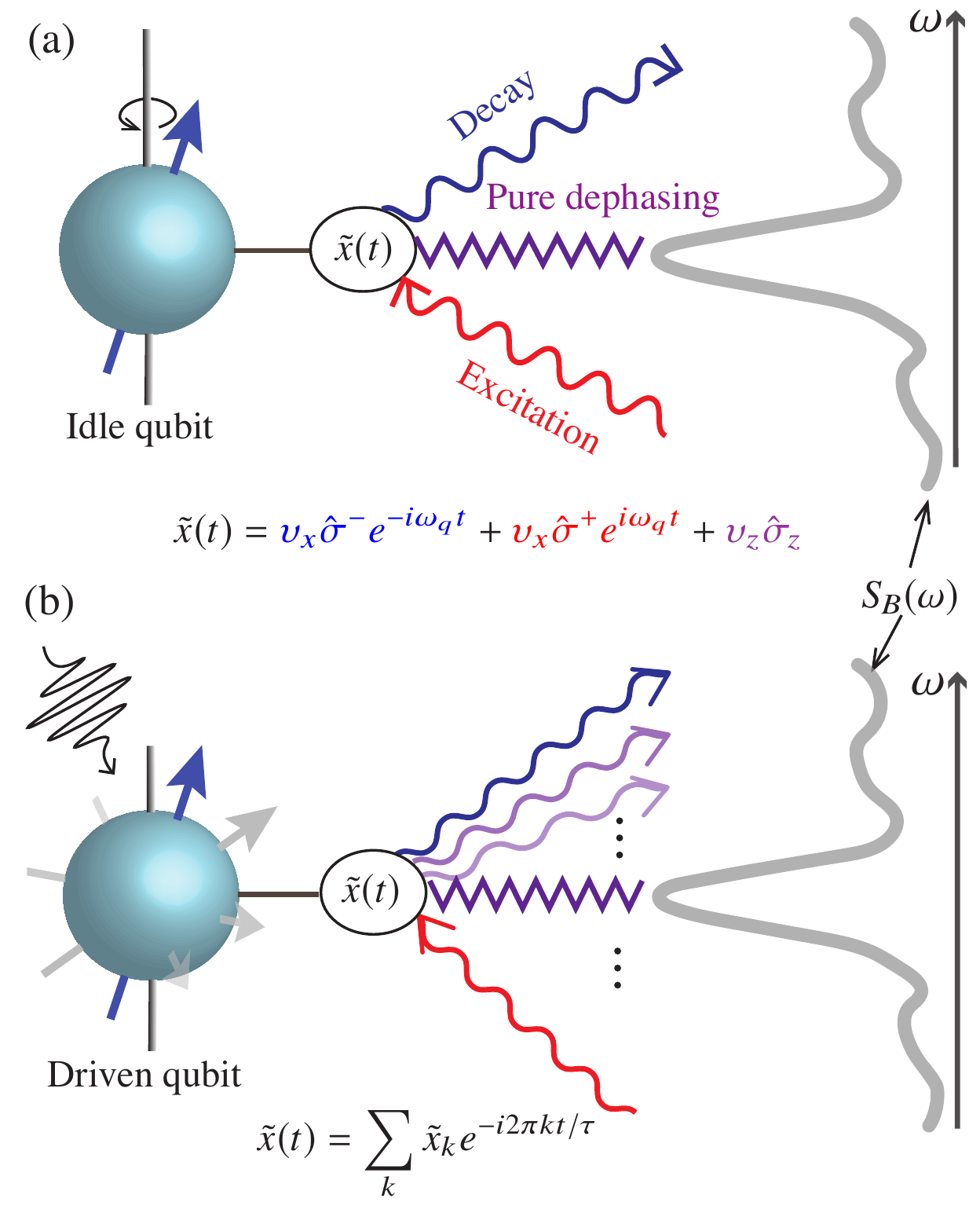}
    \caption{A cartoon comparing the decoherence channels of an undriven and driven qubit. The qubit Hamiltonian is given by $\hat{H}_{s}(t) = \omega_q\hat{\sigma}_z/2 + \hat{H}_d(t)$, where $\hat{H}_d(t)$ describes an arbitrary drive on the qubit. This qubit is coupled to the bath via the operator $\hat{x} = \upsilon_z\hat{\sigma}_z +\upsilon_x\hat{\sigma}_x$, and the bath spectrum is denoted by $S_B(\omega)$. Panels (a) and (b) illustrate the qubit decoherence channels if the qubit is undriven or driven, respectively. In (a), the three terms in the decomposition $\tilde{x}(t) = \sum_{\pm}\upsilon_x\hat{\sigma}^{\pm}\exp(\pm i\omega_qt) + \upsilon_z\hat{\sigma}_z$ result in the damping operators $\mathbb{D}[\hat{\sigma}^{\pm}]$ and $\mathbb{D}[\hat{\sigma}_z]$, which describe the excitation, decay, and pure-dephasing channels of the qubit, respectively \cite{Breuer_open_quantum_system}. In (b), because an arbitrary drive is applied on the qubit, the previous simple decomposition of $\tilde{x}(t)$ no longer holds; instead, one can use Eq.~\eqref{eq:Fourier} to perform the frequency decomposition over the time period $t\in[0,\tau]$. As a result, the decoherence channels for a driven qubit are given by the first line of the self-energy \eqref{map}.}
    \label{fig:cartoon}
\end{figure}

We append two remarks to compare our method with several existing ones. First, although our method is not a differential equation, the derived map is reminiscent of the dynamical map generated by the Lindblad equation \cite{Lindblad,Breuer_open_quantum_system}. Specifically, the first and second lines in Eq.~\eqref{map} resemble the damping terms and  Lamb shift in the master equation, respectively. A difference is that, our map considers noise contributions from the frequency set $\{k\omega_p|k\in\mathbb{Z}\}$, while the Lindblad master equation only includes noise at system transition frequencies. For a more intuitive comparison, we illustrate the decoherence channels for an undriven and driven qubit in Fig.~\ref{fig:cartoon} (a) and (b), respectively (see a detailed description of the qubit in the caption). While three damping operators fully describe the decoherence processes in an undriven qubit according to the Lindblad master equation, more damping terms are relevant for a driven one according to Eq.~\eqref{map}. In fact, we show in Sec.\ \ref{Subsec:static} that the Lindblad map is a special case of the dynamical maps derived by our method.

Second, the form of  Eq.~\eqref{map} is also reminiscent of the coarse-grained master equation \cite{Lidar_coarse_grain,Lidar_coarse_grain_drive}.  We understand this similarity as follows: the second-order secular Keldysh expansion used here is comparable to the coarse-graining step detailed in Refs.~\cite{Lidar_coarse_grain,Lidar_coarse_grain_drive}. Differently, our framework focuses on a single map of the reduced density matrix from $t=0$ to $\tau$, rather than its full evolution during $t\in [0,\tau]$. The calculation of the Fourier series of $\tilde{x}(t)$ can also be easily carried out numerically, e.g., using the method of fast Fourier transformation. Therefore, our model tends to require less computational resource due to fewer computational steps, if the map is only needed for one final time $\tau$.

\subsection{Total decoherence error}
\label{Subsec:totaldec}
Using the map \eqref{CPTP}, we can conveniently derive the process infidelity for a noisy quantum processor. Following Ref.~\cite{abdelhafez2019quantum},  this error is expressed as
\begin{align}
    E_{\mathrm{gate}}=1-\frac{1}{N_s^2}\tr\Big\{\mathcal{V}_\mathrm{tg}^\dagger \mathcal{V}_s(\tau) \mathbf{\Pi}(\tau)\Big\}.\label{eq:gterror}
\end{align}
Above, $N_s$ is the dimension of the system Hilbert space, $\mathcal{V}_\mathrm{tg}\equiv\hat{U}_\mathrm{tg} \!\otimes\!\hat{U}_\mathrm{tg}^\dagger$ denotes the target superoperator, where $\hat{U}_\mathrm{tg}$ is the target unitary, and  $\mathcal{V}_s(\tau)$ is the closed-system map defined by $\mathcal{V}_s(\tau)\equiv \hat{U}_s(\tau) \,\otimes\,\hat{U}_s^\dagger(\tau)$.

If we only focus on the decoherence contribution to $E_\mathrm{gate}$, we can neglect the possible coherent errors by setting $U_\mathrm{tg} = U_s(\tau)$, and reduce Eq.~\eqref{eq:gterror} to
\begin{align}
    E_{\mathrm{dh}}=&\,1-\frac{1}{N_s^2}\tr\{\mathbf{\Pi}(\tau)\}\nonumber\\
    \approx&\,\frac{1}{N_s}\sum_k \Big(\mathrm{Tr}_s\{\tilde{x}^\dagger_k\tilde{x}_k\} -\frac{1}{N_s}\big|\mathrm{Tr}_s\{\tilde{x}_k\}\big|^2\Big)\,\mathrm{Re}\{2\phi_{k,k}\}.\label{totalerror}
\end{align}
Above, we have used the leading-order approximation $\mathbf{\Pi}(\tau)\approx \mathbf{I}_s + \mathbf{\Sigma}_{\mathrm{CP}}^{(2)}(\tau)$. The trace $\mathrm{Tr}\{\cdot\}$ is for the superoperator for the system density matrix, while $\mathrm{Tr}_s\{\cdot\}$ denotes the usual trace for regular system operators\footnote{For a superoperator $\hat{x}\otimes \hat{y}^\dagger$, the two types of traces are related by $\mathrm{Tr}\{\hat{x}\otimes \hat{y}^\dagger\} = \mathrm{Tr}_s\{\hat{x}\}\mathrm{Tr}_s\{\hat{y}^\dagger\}$, where $\hat{x}$ and $\hat{y}$ are regular system operators.}. This approximation evaluates $E_{\mathrm{dh}}$ to the order $\epsilon^2$. Up to this order, only the real part of $\phi_{k,k}$ contributes. 

We interpret the sum in the second line of Eq.~\eqref{totalerror} as follows. The total decoherence error $E_{\mathrm{dh}}$ is a sum of contributions by noise from frequency bands indexed by $k$. The $k$th band has the approximate bandwidth $\sim\omega_p$ and is centered at $\omega_k$ [see the filter function in Fig.~\ref{fig:Filter_Func} (a)]. The total noise amplitude over this bandwidth is given by the integral $2\mathrm{Re}\{\phi_{k,k}\}$. The driven qubit, however, is not equally sensitive to noise from all these frequency bands -- according to Eq.~\eqref{totalerror}, we can quantify the sensitivity  by the filter strength  
\begin{align}
    M_k\equiv \mathrm{Tr}_s\big\{\tilde{x}^\dagger_k\tilde{x}_k\big\} -\frac{1}{N_s}\big|\mathrm{Tr}_s\{\tilde{x}_k\}\big|^2,\label{strength}
\end{align}
which satisfies the conservation rule $\sum_k M_k = \mathrm{Tr}_s\big\{\hat{x}^2\big\} - \big|\mathrm{Tr}_s\{\hat{x}\}\big|^2/N_s$ for a time-independent coupling operator $\hat{x}$. The conservation rule implies that, if only white noise is present, i.e., the noise spectrum $S_B(\omega) = \gamma$ is a constant over frequency, the decoherence error is 
\begin{align}
    E_{\mathrm{dh}} \approx \frac{1}{N_s}\Big(\mathrm{Tr}_s\big\{\hat{x}^2\big\} - \frac{1}{N_s}\big|\mathrm{Tr}_s\{\hat{x}\}\big|^2\Big)\gamma\tau,\label{eq:error-conserv}
\end{align}
which increases with $\tau$ but is independent of the shape of drive applied during $t\in[0,\tau]$ up to order $\epsilon^2$.

For non-Markovian noise [$S_B(\omega)$ is not a constant], however, Eq.~\eqref{eq:error-conserv} does not hold. In this case, different drives generally result in different magnitudes of $E_\mathrm{dh}$. To reduce decoherence errors, one should design pulses such that $M_k$ is suppressed where the integrated noise amplitude $\mathrm{Re}\{2\phi_{k,k}\}$ is large. In the following sections, most strategies discussed for reducing decoherence are centered around this strategy.

\section{Applications}
\label{Sec:Appl}
In this section, we demonstrate the power of our framework by a few examples from a wide range of applications. Our framework not only reproduces some of the established conclusions, but also extends the prediction to situations that have not been carefully studied by previous theories.

\subsection{Static quantum systems}
\label{Subsec:static}

We first apply our method to an undriven system, with the main purpose of reproducing the dynamical map derived by the Lindblad master equation. In this case, we set the drive $\hat{H}_d(t) = 0$ in Eq.~\eqref{eq:Hamiltonian}, which makes the system Hamiltonian $\hat{H}_s(t) = \hat{H}_{s0}$ time-independent.

Following the procedure described in Sec.\ \ref{Sec:DerKeldysh}, we first derive the interaction-picture coupling operator $\tilde{x}(t)$, and use it to find the filter operators $\tilde{x}_k$ needed in the derivation of the self-energy \eqref{map}. For the undriven system, the propagator is given by
\begin{align}
    \hat{U}_{s0}(t) = \exp[-iH_{s0}t],\label{Uq0}
\end{align}
which yields the coupling operator in the interaction picture
\begin{align}
    \tilde{x}(t) = \sum_{\omega_L\in\mathbb{F}}\tilde{x}(\omega_L)e^{-i\omega_L t}.
    \label{xtildeundriven}
\end{align}
Above, the frequencies $\omega_L$ associated with different terms  are contained in the frequency set $\mathbb{F} = \{E_j-E_{j'}\,|\,0\leq j,j\leq N_s\}$. For this expression, $N_s$ is the dimension of the system Hilbert space, and $E_j$ is the eigenenergy of the $j$th eigenstate for $\hat{H}_s$. We refer to the elements in $\mathbb{F}$ as the {\textit{transition frequencies}, and the corresponding $\tilde{x}(\omega_L)$ as the \textit{damping operator}. (Note that $\omega_L=0$ is also included in this transition-frequency set $\mathbb{F}$.)

Using Eq.~\eqref{eq:Fourier}, we obtain the $k$th filter operator
\begin{align}
    \tilde{x}_k=\frac{1}{\tau}\int_0^\tau\!dt' \tilde{x}(t')e^{ik\omega_p t'}=\sum_{\omega_L\in\mathbb{F}}Q(\omega_L,k\omega_p)\tilde{x}(\omega_L),
    \label{eq:xkundriven}
\end{align}
where we define $Q(\omega_L,\omega) \equiv [e^{i(\omega-\omega_L)\tau}-1]/[i(\omega-\omega_L)\tau]$. Inserting the expansion \eqref{eq:xkundriven} into Eq.~\eqref{map}, we obtain the self-energy for the undriven system as
\begin{align}
 \mathbf{\Sigma}^{(2)}_{\mathrm{CP}}(\tau) = &\sum_{k}\mathrm{Re}\{2\phi_{k,k}\}\mathbb{D}\Big[\!\! \sum_{\omega_L\in\mathbb{F}}\!Q(\omega_L,k\omega_p)\tilde{x}(\omega_L)\Big]\nonumber\\
 &+ \text{Lamb shifts},
\end{align}
where we define the damping operator $\mathbb{D}[\hat{L}]\tilde{\rho}\equiv \hat{L}\tilde{\rho}\hat{L}^\dagger -[\hat{L}^\dagger\hat{L} \tilde{\rho}+ \tilde{\rho}\hat{L}^\dagger\hat{L}]/2$. Inserting it into Eq.~\eqref{CPTP}, we obtain the dynamical map for the undriven system. In deriving this map, we only specify a time-independent Hamiltonian, but do not make further assumptions such as those usually required by the Lindblad master equation.

If we do enforce these assumptions, then our framework reproduces the Lindblad dynamical map. In detail, these conditions are: 
\begin{enumerate}

\item[\ding{172}] the difference between the transition frequencies is much larger than $\omega_p$, i.e., $|\omega_L-\omega_L'|\gg\omega_p$ for $\omega_L,\omega'_{L}\in\mathbb{F}$, $\omega_L\neq\omega'_{L}$.

\item[\ding{173}]  the system evolution time $\tau$ is sufficiently long such that the spectral variation in $S_B(\omega)$ is negligible over the small frequency scale $\omega_p=2\pi/\tau$; 
\end{enumerate}
These two conditions can be translated to the following two more familiar statements: \ding{172} the system's characteristic time $\tau_S\sim 1/\mathrm{min}\{\omega_L-\omega_L'|\omega_L,\omega_L'\in\mathbb{F}, \omega_L\neq\omega_L'\}$ is much shorter than the system evolution time $\tau$ of interest, which usually has a similar timescale as the system relaxation time $\tau_R$;  \ding{173} the bath correlation time $\tau_B$ is also much shorter than the evolution time $\tau\sim \tau_R$ (see Appendix \ref{ap:correlation} for a more detailed explanation).  Under  \ding{172} and  \ding{173}, the fundamental frequency $\omega_p$ is by far the smallest frequency scale. This allows us to consider the limit $\omega_p\rightarrow 0$, and perform the following three approximations. 

First, we approximate the function $K^R(\omega)\approx \pi\tau\delta(\omega)$ and $K^I(\omega)\approx -\tau\, \mathcal{P}({1}/{\omega})$, where $\delta(x)$ is the Dirac delta function and $\mathcal{P}$ denotes the Cauchy principal value. Using the approximated $K^{R/I}(\omega)$, we simplify the integral $\phi_{k,k}$ and obtain
\begin{align}
    \mathrm{Re}\{2\phi_{k,k}\}\approx \tau S_B(k\omega_p),\quad
    \mathrm{Im}\{\phi_{k,k}\}\approx \tau \bar{S}_B(k\omega_p),\label{reimapprox}
\end{align}
where we define 
\begin{align}
    \bar{S}_B(\omega)\equiv - \mathcal{P}\int_{-\infty}^{\infty}\frac{d\omega'}{2\pi}\frac{S_B(\omega')}{\omega'-\omega}.
\end{align}

Second, the infinitesimal $\omega_p$ also justifies the replacement of the summation over $k$ by an integral over $\omega$. This step transforms the self-energy in Eq.~\eqref{map} to an integral
\begin{align}
    \mathbf{\Sigma}^{(2)}_{\mathrm{CP}}\Tilde{\rho}_s(0)\! =\! \int_{-\infty}^{\infty}{d\omega} \Big\{& S_B(\omega) \mathbb{D}[\bar{x}_\omega]\bar{\rho}_s(0) \nonumber\\
    &-i\bar{S}_B(\omega)[\bar{x}^\dagger_{\omega}\bar{x}_\omega, \Tilde{\rho}_s(0)] \Big\},\label{mapintegral}
\end{align}
where we define $\bar{x}_{\omega}\equiv  \tilde{x}_{\lfloor\omega/\omega_p\rfloor}/\sqrt{\omega_p}$.

Third, using the expansion \eqref{xtildeundriven} for the undriven system and the definition of $\Bar{x}_{\omega}$, we find the approximation
\begin{align}
    \Bar{x}_\omega^\dagger \Bar{x}_\omega \approx&\, \frac{\tau}{2\pi} \sum_{\omega_L}\sum_{\omega_L'}Q^*(\omega_L,\omega)Q(\omega'_L,\omega)\tilde{x}^\dagger(\omega_L)\tilde{x}(\omega'_L)\nonumber\\
     \approx &\, \sum_{\omega_L} \Tilde{x}^\dagger(\omega_L) \Tilde{x}(\omega_L) \delta(\omega-\omega_L).
\end{align}
A similar delta-function approximation holds for $\bar{x}^\dagger_{\omega} \otimes \bar{x}_\omega$ in $\mathbb{D}[\bar{x}_\omega]$. Inserting these approximations into the integral Eq.~\eqref{mapintegral} and carrying out the integral over the delta functions, we finally arrive at the self-energy 
\begin{align}
    \mathbf{\Sigma}^{(2)}_{\mathrm{CP}}(\tau)\Tilde{\rho}_s(0)\! = \tau\!\! \sum_{\omega_L\in\mathbb{F}} \Big\{& S_B(\omega_L) \mathbb{D}[\tilde{x}(\omega_L)]\bar{\rho}_s(0) \nonumber\\
    &-i\bar{S}_B(\omega_L)\Big[\tilde{x}^\dagger(\omega_L)\tilde{x}(\omega_L), \Tilde{\rho}_s(0)\Big] \Big\}.\label{lindblad}
\end{align}
The secular CPTP map \eqref{CPTP} generated by the self-energy above is identical to that predicted by the Lindblad master equation. 

As a minimal example, we apply the map above to a static qubit that is described by the Hamiltonian 
\begin{align}
    \hat{H}_{s0} = \frac{1}{2}\omega_{q}\hat{\sigma}_z,
\end{align}
where $\omega_q$ is the qubit frequency. The transition-frequency set for this qubit is $\mathbb{F} = \{0,\pm\omega_q\}$. If it is transversely coupled to a bath through the operator $\hat{x} = \hat{\sigma}_x$, the expansion \eqref{xtildeundriven} is given by
\begin{align}
    \tilde{x}(t) = \hat{\sigma}^- e^{-i\omega_qt} + \hat{\sigma}^+ e^{i\omega_qt}.\label{eq:twolevelundriven}
\end{align}
Then, following the steps described above, we find the self-energy under conditions \ding{172} and \ding{173} approximated as
\begin{align}
    \mathbf{\Sigma}^{(2)}_{\mathrm{CP}}(\tau)\Tilde{\rho}_s(0) \approx \tau\sum_{\pm} S_B(\pm\omega_q) \mathbb{D}[\hat{\sigma}^{\mp}]\tilde{\rho}_s(0)+\text{Lamb shifts}.\label{lindbladqubit}
\end{align}
If the coupling operator is $\hat{x} = \hat{\sigma}_z$, we instead have  $\Tilde{x}(t)=\hat{\sigma}_z$, which yields the approximated self-energy
\begin{align}
    \mathbf{\Sigma}^{(2)}_{\mathrm{CP}}(\tau)\Tilde{\rho}_s(0) \approx \tau S_B(0) \mathbb{D}[\hat{\sigma}_z]\tilde{\rho}_s(0)+\text{Lamb shifts}.\label{lindbladqubitz}
\end{align}

\subsection{Weakly driven systems}
\label{Subsec:weak}
Although the derivation of the Lindblad dynamical map \cite{Breuer_open_quantum_system} assumes no drive on the system, in the literature, weak drives are sometimes naively added to the master equation with the damping rates and operators unaffected. Different from the Lindblad method, our framework rigorously includes the drive in the derivation. Below, we use the Keldysh framework to investigate the change of the decoherence map \eqref{CPTP} if a weak drive is added. Because the specific noise spectrum may vary in different experiments, here we choose to focus on the filter operators $\Tilde{x}_k$, which determine the map \eqref{CPTP} up to the specific noise spectrum. 

We start by considering a general system, which is described by the Hamiltonian $\hat{H}_s(t) = \hat{H}_{s0} + \lambda\hat{H}_d(t)$. This Hamiltonian consists of the static part $\hat{H}_{s0}$ and a sufficiently weak driving term $\lambda\hat{H}_d(t)$ ($\lambda$ is a small dimensionless parameter). Due to the small amplitude of the latter term, we can perturbatively calculate $\hat{U}_s(t)$ using the Magnus expansion \cite{BLANESMagnus}:
\begin{align}
    \hat{U}_s(t) = \hat{U}_{s0}(t)\hat{U}_d(t),\quad \hat{U}_d(t)=\exp[-i\hat{\Omega}(t)],\label{eq:uqud}
\end{align}
where $\hat{U}_{s0}(t)$ is the propagator for the undriven system given in Eq.~\eqref{Uq0}, and $\hat{\Omega}(t)$ is the Magnus exponent. To leading order of $\lambda$, this exponent is approximately
\begin{align}
    \hat{\Omega}(t)\approx \lambda\int_0^t dt' \hat{U}^\dagger_{s0}(t')\hat{H}_d(t')\hat{U}_{s0}(t').
\end{align}

By inspecting Eq.~\eqref{eq:uqud} and the definition of $\tilde{x}(t)$, we note that: if $\hat{\Omega}(t)$ is small, $\tilde{x}(t)$ can be approximated as
\begin{align}
    \Tilde{x}(t) \approx &\, \sum_{\omega_L\in\mathbb{F}}\Big\{\tilde{x}(\omega_L)-i[\tilde{x}(\omega_L), \hat{\Omega}(t)]\Big\}e^{-i\omega_L t},\label{xtildedrivenapprox}
\end{align}
which is only slightly modified from Eq.~\eqref{xtildeundriven}. In the limit $|\hat{\Omega}(t)|\rightarrow 0$, the filter operator $\Tilde{x}_k$ can still be approximated by the undriven expansion \eqref{eq:xkundriven}. Therefore, the noise channels and the resulting dynamical map \eqref{CPTP} should also approach those obtained in the undriven case. By contrast, if the condition of negligible $\hat{\Omega}(t)$ is not satisfied, such approximation may be invalid. 

In the following, we will use a concrete example to concretely demonstrate both scenarios. We consider a qubit driven by a sinusoidal tone. The Hamiltonian of this driven system is given by 
\begin{align}
    \hat{H}_{s0} = \frac{\omega_q}{2}\hat{\sigma}_z,\quad \hat{H}_d(t) = \frac{d}{2}(\hat{\sigma}^+e^{-i\omega_d t} + \hat{\sigma}^-e^{i\omega_d t}).\label{eq:qubit_Hamiltonian}
\end{align}
 The coupling operator for this qubit is taken to be $\hat{x} = \hat{\sigma}_x$, which corresponds to the transverse coupling between the qubit and the noise bath. The drive strength $d$ is assumed to be weak, i.e., $d\ll \omega_q$. In that case, the Magnus expansion of the qubit propagator is applicable, with the exponent given by
 \begin{align}
     \hat{\Omega}(t) =  \frac{d}{2\delta_q}\,\mathrm{sin}(\delta_qt)   \hat{\sigma}_x +  \frac{d}{\delta_q}\mathrm{sin}^2\left(\frac{\delta_qt}{2}\right)\hat{\sigma}_y + O(d^2),\label{Omega_exp}
\end{align}
 where the detuning is defined by $\delta_q\equiv\omega_q-\omega_d$.

From Eq.~\eqref{Omega_exp}, we find that the Magnus exponent has a negligible magnitude, i.e., $|\hat{\Omega}(t)|\lesssim 4|d/\delta_q|$, if the drive is off-resonant ($d\ll |\delta_q|$). For such off-resonant drive, Eq.~\eqref{xtildedrivenapprox} predicts that the expansion of $\tilde{x}_k$ can be approximated by Eq.~\eqref{eq:twolevelundriven}. To verify this, we numerically \cite{qutip} calculate $\tilde{x}_k$  and the resulting filter strength $M_k$ for both undriven  (red coloring) and off-resonantly driven (blue coloring) qubits. The resulting filter strengths $M_k$ versus filter frequencies $\omega =k\omega_p$ are shown in Fig.~\ref{fig:FilterWeights} (a). [We only focus on the frequency range $\omega\approx \omega_q$ as an example, where noise induces energy decay in the qubit. For qubit excitation, the discussion is analogous.]  As shown in the plot, the filter strengths for the driven qubit only insignificantly differ from those for the undriven qubit. The filter operator associated with the most prominent filter strength is approximately $\hat{\sigma}^-$, which is the decay operator for the undriven qubit (red peak). 
 
For the resonantly-driven qubit, however, the exponent $\hat{\Omega}(t)$ can grow significantly, even in the limit $d\ll \omega_q$. Choosing $\delta_q = 0$ as an example, we find that the exponent
\begin{align}
    \hat{\Omega}(t) = \frac{d}{2}t\hat{\sigma}_x,
\end{align}
 grows linearly with time. This exponent leads to the expression for the coupling operator
\begin{align}
    \tilde{x}(t) =&\, \hat{U}^{\dagger}_s(t)[\hat{\sigma}^+ e^{i\omega_qt} + \hat{\sigma}^- e^{-i\omega_qt}]\hat{U}_s(t)\label{xtdrivenres}\\
    =&\,\hat{\sigma}_x\cos(\omega_qt)-[\hat{\sigma}_y\cos(dt) - \hat{\sigma}_z\sin(dt)]\sin(\omega_qt).\nonumber
\end{align}

Certainly, Eq.~\eqref{xtdrivenres} cannot be approximated by  Eq.~\eqref{xtildeundriven}, rendering the previous approximation of $\tilde{x}_k$ by Eq.~\eqref{eq:xkundriven} invalid. The difference in $\tilde{x}_k$ between the driven and undriven cases causes distinctive behaviors of $M_k$'s, as shown in Fig.~\ref{fig:FilterWeights} (b).  Compared to the plot of $M_k$ for the undriven qubit  (red coloring), the plot for the driven qubit (blue coloring) exhibits two additional peaks located at frequencies $\omega=\omega_q\pm d$. These extra peaks imply additional decoherence channels, rendering the qubit sensitive to noise at frequencies $\omega_q\pm d$ in addition to its transition frequency $\omega_q$. These additional damping channels are missed by the standard Lindblad method [Eq.~\eqref{lindbladqubit}], but can be crosschecked by a Golden-rule type of calculation in the rotating frame \cite{Oliver_rotating_frame_relaxation}, or the Floquet theory \cite{Dynamical_sweet_spot}. (In Appendix \ref{floquet_weak}, we explain the appearance of the side peaks in the framework of the Floquet master equation.) In summary, the filter operators $\tilde{x}_k$ for the driven system in general differ from those in the undriven case. This further results in different dynamical maps \eqref{CPTP}. The approximation of one set of $\tilde{x}_k$ by the other is possible if the drive is off-resonant such that $|\hat{\Omega}(t)|$ is sufficiently small\footnote{If only white noise is present, however, the total decoherence errors in the driven and undriven cases equal. See discussion around Eq.~\eqref{eq:error-conserv}.}.

Finally, we use a third example to demonstrate the predictive power of our method in more complicated situations involving non-periodic drives. For example, we calculate $\tilde{x}_k$ and $M_k$ for a qubit driven by a pulse with a hyperbolic envelope [see inset of Fig.\ \ref{fig:FilterWeights} (c)].  Compared to the plot of filter strengths in (b), the ramping up and down of the drive result in two wider side peaks that are not centered at the maximal driving strength, as shown in (c). Such distinctive feature implies the difference in the decoherence processes between systems with periodic and non-periodic drives. The latter case is thought to go beyond the description by the rotating-frame analysis or the Floquet theory, but is conveniently captured by the Keldysh method. 

\begin{figure}[h!]
    \centering
\includegraphics[width=6.5cm]{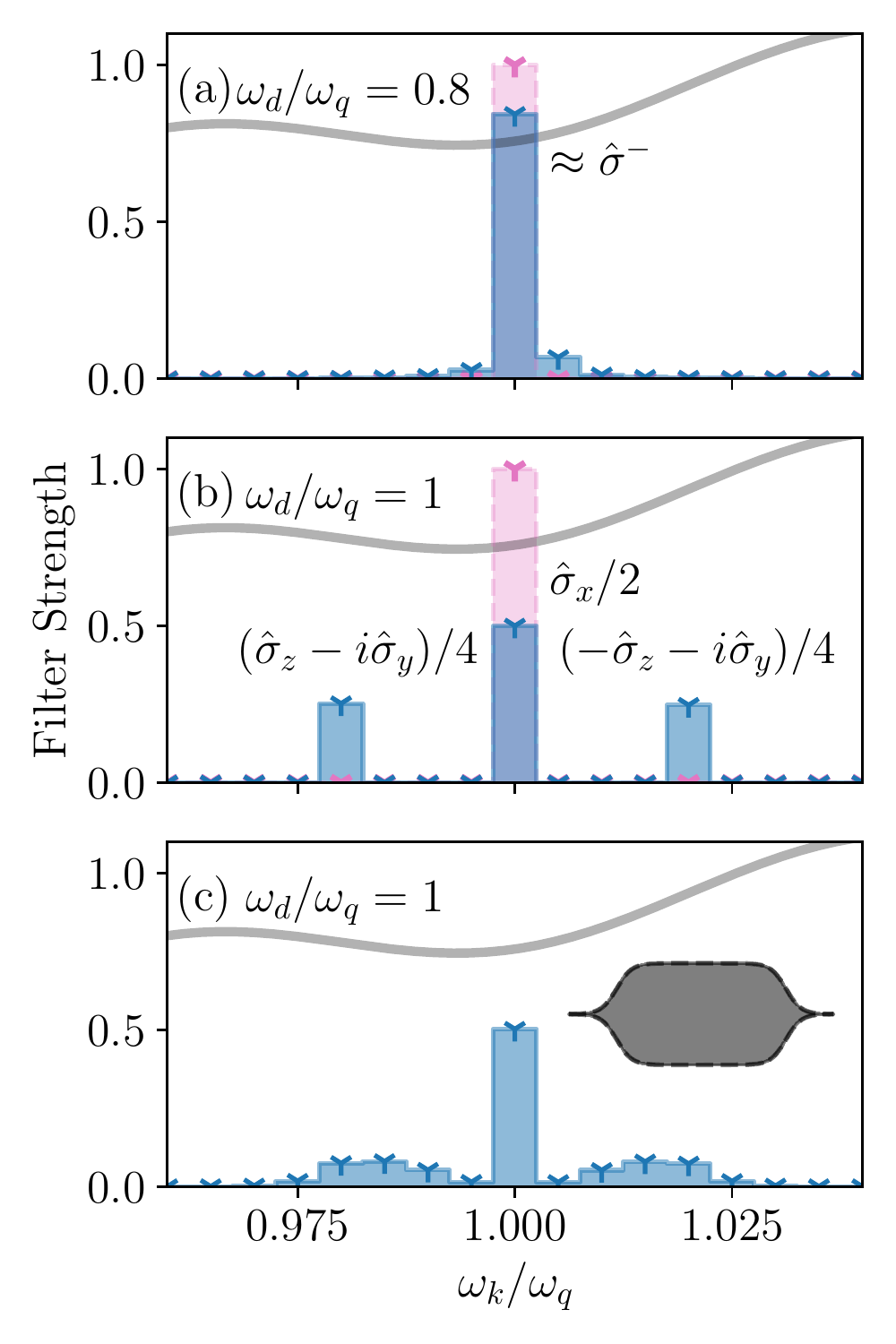}
    \caption{Filter strength $M_k$ for a driven qubit described by Hamiltonian Eq.~\eqref{eq:qubit_Hamiltonian}. In (a)-(c), the horizontal axis shows the filter frequency $\omega_k=k\omega_p$, and the vertical shows the filter strength $M_k$ [Eq.~\eqref{strength}]. The widths of the columns are given by the fundamental frequency $\omega_p=2\pi/\tau$. The parameters are chosen as follows. The drive amplitudes for all three simulations are chosen as $d/\omega_q = 0.02$, and the frequency $\omega_d$ used for each plot is given in each figure. The duration is set as $\tau = 200\cdot 2\pi/\omega_q$ for all three simulations. For (a) and (b), we choose sinusoidal drives with a constant amplitude; the results of $M_k$ for the driven qubit are shown in blue, while those for an undriven qubit are shown in red for reference. For (c), the sinusoidal drive used for (b) is multiplied by a hyperbolic envelope $\mathcal{F}(t) = [1+\mathrm{tanh}[(t-t_{\mathrm{mid1}})/\pi t_\mathrm{ramp}]][1+\mathrm{tanh}[(t_{\mathrm{mid2}}-t)/\pi t_\mathrm{ramp}]]/4$, as shown in the inset. ($t_\mathrm{mid1,2}$ and $t_{\mathrm{ramp}}$ are used to tune the location and duration of the ramps.)}
    \label{fig:FilterWeights}
\end{figure}

\subsection{Ramsey, echo and 1/$f$ noise}
\label{Subsec:1/f}

Along with the introduction of Eq.~\eqref{map} in Sec.\ \ref{Subsec:Fourierexp}, we claim that the secular approximation is applicable even for spectra that show strong variation within the frequency scale characterized by $\omega_p$. Here, as a supporting example, we study the state evolution of a qubit which is coupled to a $1/f$ noise source, whose spectrum is strongly peaked at $\omega\approx 0$. Particularly, we compare the prediction by the secular CPTP map \eqref{CPTP} and the full-wave version \eqref{map_full} for this example.

We consider a qubit longitudinally coupled to a noise bath and subject to a transverse drive. The  Hamiltonian for this qubit is $\hat{H}_s(t) = \omega_q\hat{\sigma}_z/2+d(t)\hat{\sigma}_x$, and the coupling operator is $\hat{x} = \hat{\sigma}_z$.  The $1/f$ noise spectrum is given by $S_B(\omega) = 2\pi \mathcal{A}_f^2/|\omega|$, where we set an infrared cutoff frequency $\omega_{\mathrm{ir}}$ to regularize the singularity at $\omega=0$.

We first calculate the map \eqref{CPTP} for the simple case of an undriven qubit [$d(t)=0$], which is relevant for a Ramsey experiment \cite{Ithier_decoherence_analysis}. Different from the discussion in Sec.\ \ref{Subsec:static}, the presence of the strong peak in the noise spectrum violates condition \ding{173}, rendering the Lindblad prediction \eqref{lindbladqubitz} invalid. This difficulty, instead, can be overcome by our Keldysh framework, which takes advantage of the filter functions \cite{Green_filer_func}. To perform the Keldysh calculation, we first derive the filter operator for the undriven qubit $\Tilde{x}_k = \delta_{k,0}\hat{\sigma}_z$, meaning that $\Tilde{x}_{k=0}$ is the only non-vanishing filter operator. Such decomposition enables the analytical evaluation of both Eqs.~\eqref{map} and \eqref{map_full}, which predicts the same self-energy exponent
\begin{align}
    \mathbf{\Sigma}^{(2)}(\tau)=&\,\int^{\infty}_{-\infty} \frac{d\omega}{2\pi}S_B(\omega)2K^R(\omega)\mathbb{D}[\hat{\sigma}_z]\nonumber\\
    \approx&\, 2\mathcal{A}_f^2\tau^2 \ln\left(\frac{1}{2\pi\omega_\mathrm{ir}\tau}\right)\mathbb{D}[\hat{\sigma}_z].\label{1/f}
\end{align}
The log-quadratic scaling of the self-energy implies a well-known sub-Gaussian dephasing profile \cite{Ithier_decoherence_analysis, Fluxonium_hc}, which differs from the exponential one predicted by the Lindblad map \eqref{lindbladqubitz}.

\begin{figure}
    \centering
\includegraphics[width=7.5cm]{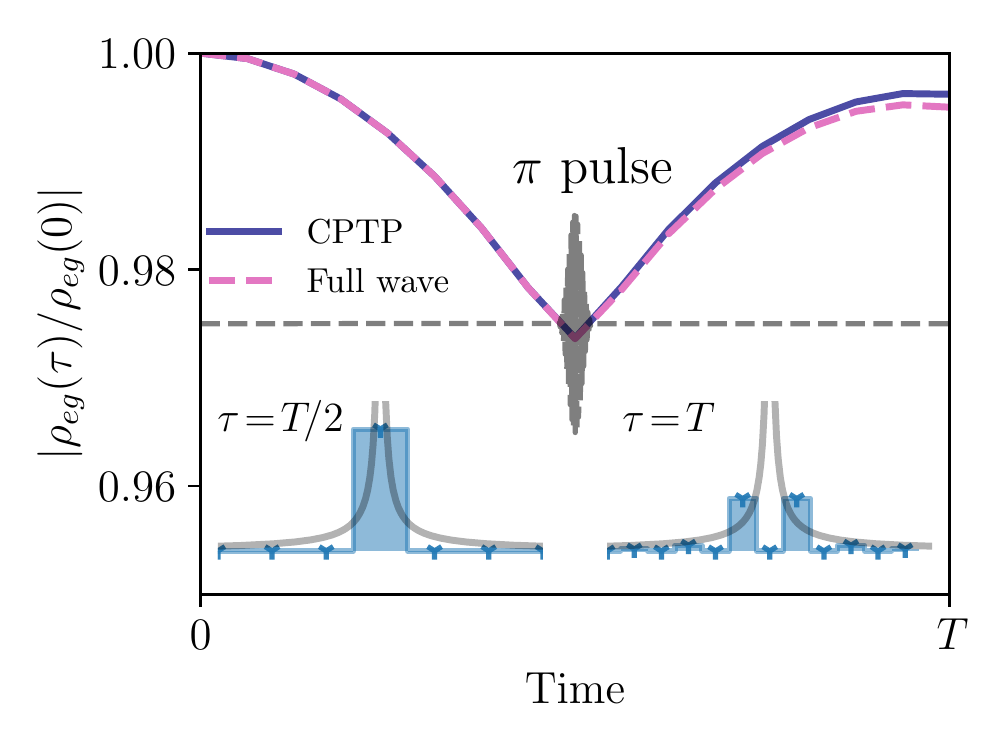}
    \caption{Evolution of $\tilde{\rho}_{eg}(\tau)$ for a qubit during a spin echo experiment. The blue solid and red dashed curves correspond to the predictions by Eq.~\eqref{CPTP} and Eq.~\eqref{fullwave}, respectively. The sketch of the echo pulse used for this simulation is plotted by the gray dashed curve. The two insets show the filter strengths $M_k$ as functions of the filter frequencies $\omega_k$, at the middle and end of the echo duration. Along with the filter strengths, the $1/f$ noise spectrum used for simulation is also sketched.}
    \label{fig:Echo}
\end{figure}

As the secular and full-wave maps agree well for the undriven qubit, we next check whether that agreement extends to the driven case. We focus on a qubit undergoing a spin echo. This protocol uses a $\pi$-pulse to help refocus the phase of the qubit and, as a result,  mitigate the qubit dephasing due to $1/f$ noise. 

For this simulation,  we choose the drive such that a finite-width $\pi$ pulse is applied at the middle of the whole echo duration $T$ (gray curve in Fig.~\ref{fig:Echo}). Due to the application of the pulse, the filter strengths for different $\tau\in[0,T]$ differ characteristically. For example, we calculate $\Tilde{x}_k$ for $\tau=T/2$ and $\tau=T$, and show the corresponding filter strength $M_k$ in the insets of Fig.~\ref{fig:Echo}. For $\tau=T$, the filter strength $M_{k=0}$ vanishes, and the most prominent peaks of $M_k$ are located at $k=\pm 1$. By contrast, for $\tau=T/2$, the only prominent filter strength is $M_{k=0}$, indicating strong sensitivity to noise from $\omega\approx 0$. 

Using the filter operators and the $1/f$ noise spectrum, we further calculate the decoherence maps predicted by both Eq.~\eqref{fullwave} and Eq.~\eqref{CPTP}, with $\tau$ varied over $\tau\in[0,T]$. Note that since this framework is not based on a differential equation, each $\Tilde{\rho}_s(\tau)$ with $\tau\in[0,T]$ is calculated separately rather than recursively. Then, we evaluate the off-diagonal matrix element $\Tilde{\rho}_{eg}(\tau)$, which is plotted in Fig.~\ref{fig:Echo}. The magnitude of this matrix element indicates the phase coherence of the qubit \cite{Groszkowski_Zero_pi_theory}, if the qubit is initially prepared in an equal superposition state. 

Visibly, the two calculations show qualitative agreement, while a small deviation exists as the consequence of neglecting the off-diagonal filter functions in Eq.~\eqref{map}. The comparison suggests that, even for a non-trivial decomposition $\Tilde{x}_k$ and a highly structured noise spectrum, the secular approximation can still be applicable\footnote{For numerical simplicity, we can always choose to truncate the expansion in Eq.~\eqref{map_full} by only keeping terms with $|k-k'|$ smaller than a certain integer, since those with a larger $|k-k'|$ are less important due to their diminishing amplitude according to Eq.~\eqref{ampdecay}. But one should be alerted that such calculation has the risk of generating a non-CPTP map.}. 

Besides the agreement between the two sets of simulations, we also observe the interesting rebound of the matrix element $|\tilde{\rho}_{eg}(\tau)|$. Specifically,  $|\tilde{\rho}_{eg}(\tau)|$ decreases during the first half of the echo period and then increases after the pulse is applied. [For the first half period, the evolution of $\tilde{\rho}_{eg}(\tau)$ indeed shows the sub-Gaussian dephasing behavior predicted by Eq.~\eqref{1/f}.] Such ``inverse dephasing" of the qubit implies a negative decoherence rate, which is described in more detail by a time-local master equation introduced in Ref.~\cite{Groszkowski_Magnus_master}. This behavior also sends another useful message: although the map $\mathbf{\Pi}(\tau) \approx \exp[\mathbf{\Sigma}_{\mathrm{CP}}^{(2)}(\tau)]$ is guaranteed to be CPTP, the intermediate map $\mathbf{\Pi}(\tau)[\mathbf{\Pi}(\tau')]^{-1}$ ($0<\tau'<\tau$) is not necessarily so. Noticing this, one may naturally ask whether it is possible to follow the procedure in Sec.\ \ref{Sec:DerKeldysh} to derive a CPTP map from $t=\tau'$ to $\tau$ ($0<
\tau'<\tau$), if the system and bath are initialized at $t=0$; however, we point out that the basis for such derivation may not hold -- for $t=\tau'>0$, the two subsystems may already be entangled, while the derivation starting from Eq.~\eqref{eq:rhoqt} requires isolation between them. (See discussion of the relation between initial entanglement and the CPTP character of the map in Ref.~\cite{Rodriguez-Rosario_CPTP}.)

\subsection{Floquet qubits}
\label{Subsec:floquet}

We designate this final subsection to test the Keldysh method in studying the Floquet qubit \cite{Dynamical_sweet_spot,Dynamical_sweet_spot_exp,Blais_Floquet,Siddiqi_floquet_qubit,Trif_spin_qubit_floquet}. This type of qubit uses the Floquet states of a periodically driven system to store and manipulate quantum information, which can offer advantages such as increased coherence times and more convenient gate operations than the static qubits. Although the open-system Floquet theory is developed to calculate the decoherence rates in such systems, its applicability is limited to the idle Floquet qubit. In the following, we show that the Keldysh method not only reproduces some results by such theory, but also explores situations that are beyond its application.

We start by studying an idle Floquet qubit, where the drive $\hat{H}_d(t)$ in Eq.~\eqref{eq:Hamiltonian} is periodic, i.e., $\hat{H}_d(t+T_d) = \hat{H}_d(t)$ ($T_d=2\pi/\omega_d$ is the drive period). In this case, the closed-system propagator can be expressed as
\begin{align}
    \hat{U}_s(t) = \sum_j\vert w_j(t)\rangle\langle w_j(0)\vert e^{-i\varepsilon_jt},
\end{align}
where $\vert w_j(t)\rangle$ and $\varepsilon_j$ are the $j$th independent Floquet state and its corresponding quasi-energy. In the interaction picture, the coupling operator is transformed as
\begin{align}
    \tilde{x}(t) = &\, \sum_{j,j'} \vert w_j(0)\rangle\langle w_{j'}(0)\vert\times\langle w_j(t)\vert \hat{x}\vert w_{j'}(t)\rangle e^{-i(\varepsilon_{j'}-\varepsilon_{j})t} \nonumber\\
    =&\sum_{\omega_L\in\mathbb{F}}\hat{x}(\omega_L)e^{-i\omega_L t},
\label{eq:floquet_xtilde}
\end{align}
where the set $\mathbb{F} = \{\varepsilon_j-\varepsilon_{j'}+l\omega_d\,|\,0<j,j\leq N_s, l\in\mathbb{Z}\}$ contains all possible quasi-energy differences; the operators $\hat{x}(\omega_L)$ are damping operators in the basis of Floquet states (time-independent in the interaction picture), rather than the eigenstates of the undriven qubit discussed in Sec.\ \ref{Subsec:static}.

With these preparations, we next show how our Keldysh method reproduces the prediction of the decoherence process via the Floquet theory in Ref.~\cite{Dynamical_sweet_spot}. In that instance, a Floquet qubit is coupled to both the $1/f$ flux-noise bath and dielectric noise bath. The spectrum of the former has a strong peak at $\omega\approx 0$, and that of the latter has a smoother spectrum.

Similar to the steps in that reference, we first disregard the peak at $\omega=0$ in the spectrum, and then correct the resulting dynamical map with a more careful consideration of the peak. For the first step, both conditions \ding{172} and \ding{173} are satisfied if we assume a sufficiently large evolution time $\tau$. If so, the map takes on the same form as Eq.~\eqref{lindblad}, while the operators $\tilde{x}(\omega_L)$ are updated according to Eq.~\eqref{eq:floquet_xtilde}. Such map reproduces that generated by the Markovian Floquet master equation \cite{Breuer_open_quantum_system}. Then, to address the strong peak at $\omega=0$ due to the $1/f$ spectrum, we carefully evaluate the coefficient $\mathrm{Re}\{2\phi_{0,0}\} = \int_{-\infty}^{\infty} (d\omega/2\pi) 2K^R(\omega)S_B(\omega)$, which is found to be approximately $2\mathcal{A}^2_f\tau^2\ln|2\pi\omega_{\mathrm{ir}}|$. This coefficient should replace $\tau S_B(\omega_L=0)$ multiplying the damping term $\mathbb{D}[\Tilde{x}(\omega_L = 0)]$  in Eq.~\eqref{lindblad}. After these two steps, the decoherence map derived in Ref.~\cite{Dynamical_sweet_spot} is reproduced exactly.

\begin{figure}[h!]
    \centering
\includegraphics[width=7.5cm]{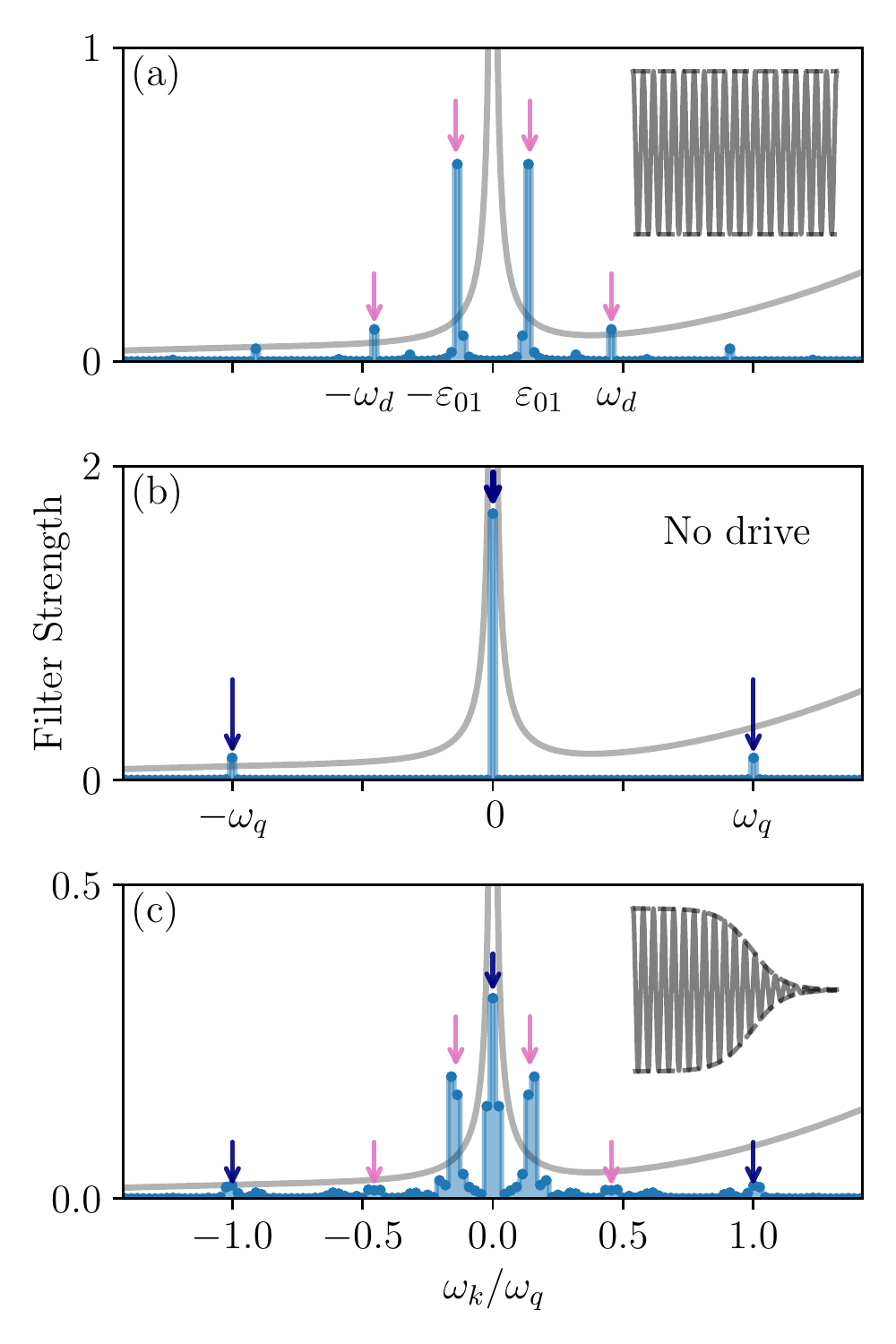}
    \caption{Filter strength $M_k$ for a Floquet qubit \cite{Dynamical_sweet_spot}. The qubit Hamiltonian is given by $\hat{H}_s(t) = [{\Delta}\hat{\sigma}_x +(2A\cos\omega_{d}t+{B})\hat{\sigma}_z)]/2$, where the parameters are chosen as $\omega_d/\Delta = 1.17$ and $B/\Delta = 1.37$. The static qubit frequency is $\omega_q = \sqrt{\Delta^2 + B^2
    }$. The evolution time is set as $\tau = 20\cdot 2\pi/\omega_d$ for all three plots. For (a), we set $A/\Delta = 2.27$. The filter strength $M_{k=0}$ vanishes, which corresponds to a dynamical sweet spot. For (b), the qubit is undriven ($A=0$). In this case, the filter strength $M_{k=0}$ is predominant, implying strong sensitivity to 1/$f$ noise. For (c), the drive used in (a) is continuously switched off according to a hyperbolic envelope $\mathcal{F}(t) = [1+\mathrm{tanh}[(t_{\mathrm{mid}}-t)/\pi t_\mathrm{ramp}]]/2$, as shown in the inset ($t_\mathrm{mid}$ and $t_{\mathrm{ramp}}$ are used to tune the starting time and duration of the switch-off, respectively). The resulting plot of $M_k$ differs from those in both (a) and (b). In (a)-(c), the noise spectra assumed in Ref.~\cite{Dynamical_sweet_spot} is also plotted in gray.}
    \label{fig:FloquetFilter}
\end{figure}

The discussion above focuses on an idle Floquet qubit. For gate operations and readout on a Floquet qubit, non-periodic control is required, and the Floquet theory is no longer applicable. Remarkably, the Keldysh method is still useful in predicting the decoherence map, since the knowledge of the Floquet states and their quasi-energies are not prerequisites for our numerical calculation of the Fourier expansion of $\tilde{x}(t)$. Below, we show one such example, where a Floquet qubit undergoes an adiabatic evolution from a dynamical sweet spot to an unprotected static working point. (This previously has been used for readout of Floquet qubits \cite{Lupascu_flux_qubit_Floquet,Dynamical_sweet_spot_exp,Siddiqi_floquet_qubit}.)

For a concrete simulation, we reuse the qubit model and parameters in Ref.~\cite{Dynamical_sweet_spot} [Fig.~3 (b) and (c) of that reference]. Specifically, we consider a fluxonium qubit with its external flux $\phi_e$ biased slightly away from the half-flux-quantum sweet spot. Under a periodic drive that is carefully tuned, the qubit can be operated at a so-called dynamical sweet spot, where the derivative $\partial \varepsilon_{01}/\partial \phi_e$ vanishes. This leads to the first-order insensitivity of the qubit to the $1/f$ flux noise. (See more details in the caption of Fig.~\ref{fig:FloquetFilter} in the current paper.) 

As references, we first calculate the filter operators $\tilde{x}_k$ for the dynamical sweet spot (initial) and the static point (final), and plot the resulting $M_k$ in (a) and (b), respectively. At the dynamical sweet spot, the qubit is to the first order insensitive to $1/f$ noise, as shown by the vanishing filter strength $M_{k=0}$ with the corresponding filter frequency $\omega_k=0$. In turn, it is sensitive to noise at other Floquet transition frequencies contained in $\mathbb{F}$ [the locations of these frequencies are pointed to by pink arrows in Fig.~\ref{fig:FloquetFilter} (a)]. By contrast, the qubit at the static working point is strongly sensitive to $1/f$ noise, as indicated by the large filter strength $M_{k=0}$. In addition, the qubit is also sensitive to noise at the qubit frequencies $\pm\omega_q$. These frequencies are marked by the blue arrows in Fig.~\ref{fig:FloquetFilter} (b). We note that the plots for $M_k$ in both (a) and (b) are reminiscent of the plot for the filter weights in Ref.~\cite{Dynamical_sweet_spot} [Fig.~3 (b) and (c) of that reference]. In fact, the resulting dynamical maps reproduce those obtained in Ref.~\cite{Dynamical_sweet_spot}, as long as the evolution time $\tau$ is taken to be sufficiently large.

Finally, we calculate $M_k$ for the adiabatic process connecting the two working points, and show the results in (c). For this case, the plot of $M_k$ differs from those from both (a) and (b). Interestingly, the locations of the peaks in (c) overlap with those from both (a) and (b). [We mark the peak locations with arrows with different coloring to indicate their apparent origin.] This feature suggests that, the qubit undergoing the adiabatic evolution is subjected to a combination of decoherence channels from both the Floquet and static regimes. Such a feature cannot be predicted by the open-system Floquet theory.

The examples used in this and the previous subsections are limited to qubits with only two energy levels, while the theory developed in Sec.~\ref{Sec:DerKeldysh} applies to general quantum systems. In Appendix \ref{ap:oscillator}, we apply this framework to derive the dynamical map for an arbitrarily driven harmonic oscillator as an example.

\section{Quantum optimal control}
\label{Sec:optimal}
Above, we have introduced the secular CPTP map and applied it in studying a variety of driven systems. Besides predicting the decoherence maps, the framework can also be used to design drive pulses that mitigate decoherence errors, once it is combined with the technique of quantum optimal control \cite{koch2022quantum,gunther2021quandary,abdelhafez2019gradient,Schuster_robust_optimal_control,Norris_filter_func,Bluhm_filter_functions,Bluhm_filter_functions_PRR,Biercuk_optimal_control,qcontrol_filter_func,Du_optctrl_cnot,Bluhm_optimal_control}.

Optimal-control techniques utilize computer-aided optimization of pulses to minimize state-transfer or gate infidelities. For open-system optimization, the widely used decoherence model is the Lindblad master equation  \cite{gunther2021quandary,abdelhafez2019gradient}. Besides this model, Refs.~\cite{Norris_filter_func,Bluhm_filter_functions,Bluhm_filter_functions_PRR,Biercuk_optimal_control,qcontrol_filter_func,Du_optctrl_cnot,Bluhm_optimal_control} have used the filter-function approach to reduce the error in quantum operations induced by correlated classical noise. 
 
 In the following, we implement our Keldysh decoherence model and extend the filter-function-aided optimization to the mitigation of quantum noise. The optimization with our method is also relatively simpler. In fact, once the total evolution time $\tau$ is set, the integrals for evaluating the coefficients $\phi_{k,k}$ can be precalculated, which are independent of the form of the drives. The CPTP maps \eqref{CPTP} also avoid the risk of optimization over unphysical maps. In the following, we use two examples to showcase the capability of the Keldysh-assisted quantum optimal control.

\subsection{State transfer in the presence of Ohmic noise}
\label{Subsec:state}
We first study the state transfer in a qubit coupled to a typical quantum-noise bath, an Ohmic noise. The spectrum of such noise is $S_B(\omega) = \mathcal{A}_o\omega \,\Theta(\omega)$ [assuming zero temperature; see inset of Fig.~\ref{fig:StateTransfer} (a) for the spectrum], where $\mathcal{A}_o$ denotes the noise strength and $\Theta(\omega)$ is the Heaviside function. The qubit coupled to this bath is described by the Hamiltonian $\hat{H}_s(t) = \omega_q\hat{\sigma}_z/2+d(t)\hat{\sigma}_x$, where $d(t)$ denotes the drive field. The coupling operator is set as $\hat{x} = \hat{\sigma}_x$.

In this setup, the spectrum exhibits clear asymmetry between the amplitudes of noise at positive and negative frequencies, which implies distinctive excitation and decay rates in the idle qubit [see Eq.~\eqref{lindbladqubit}]. Because of this, if the decay rate overwhelms the excitation rate, one can leverage the natural system-bath interaction to realize the $\vert e\rangle\rightarrow\vert g\rangle$ transfer; for the reverse transfer, however, such decay should instead be carefully mitigated. We note that the usual filter-function formalism is not designed to resolve such asymmetry in the noise spectrum, which is a difficulty we can overcome using our method. 

For our example, we consider the more difficult $\vert g\rangle\rightarrow\vert e\rangle$ transfer. To mitigate the error caused by the energy decay, we program the optimizer to minimize the infidelity 
\begin{equation}
    E_{\mathrm{st}}=1-\Big|\tr\big[\hat{U}^\dagger_s(\tau)\hat{\rho}_e\hat{U}_s(\tau)\mathbf{\Pi}(\tau)\hat{\rho}_g\big]\Big|^2.\label{eq:state-transfer}
\end{equation}
Above, $\hat{U}_s(\tau)$ is the closed-system  propagator, $\mathbf{\Pi}(\tau)$ is the CPTP map \eqref{CPTP}, and the two density matrices are $\hat{\rho}_{e(g)}\equiv\vert e(g)\rangle\langle e(g)\vert$. The implemented optimization algorithm is Gradient Ascent Pulse Engineering, which is commonly used in quantum optimal control \cite{khaneja2005optimal,Leung_optimal_control}. 

For comparison, we first optimize the pulse assuming that the noise is absent. In this case, the optimizer chooses a pulse  reminiscent of a typical resonant Rabi drive [red curve in Fig.~\ref{fig:StateTransfer} (a)], whose amplitude is almost constant.  (The step-like structure of the pulse is a result of our requirement of the piecewise-constant drive.) For a closed-system simulation, the pulse induces a smooth increase of the excited-state population, resulting in a negligible state-transfer error ($<10^{-6}$) at the end of the pulse. However, for the open-system simulation including the noise bath, the calculation by Eq.~\eqref{CPTP} predicts a much higher error ($E_{\mathrm{st}}=8.8\times 10^{-2}$), which is caused by the interaction between the qubit and the Ohmic noise bath. Especially, the smooth increase in population renders the qubit prone to the energy loss for the whole state-transfer duration [red dashed curve in (b)].

We next optimize the pulse with the noise included in Eq.~\eqref{eq:state-transfer}. The optimized pulse is shown by the blue curve in Fig.~\ref{fig:StateTransfer} (a). Different from the closed-system version, the amplitude of the open-system optimized drive is held close to zero until the latter half of the duration, where the amplitude is ramped up rapidly. In this way, the qubit stays in the excited state for a shorter time than in the previous version. Such behavior of the excited-state population reduces the decoherence error [see the comparison between blue and red curves in Fig.~\ref{fig:StateTransfer} (b)], yielding a 4.4$\times$ reduction in the state-transfer infidelity ($E_{\mathrm{st}}=2.0\times 10^{-2}$).
 
We note that a similar result is obtained by Ref.~\cite{abdelhafez2019gradient}, where the optimization is based on quantum trajectories. For comparison, that optimization presumed the knowledge of the damping rates and operators, which is derived for an idle qubit using the Lindblad equation rather than a driven system [see discussion in Sec.\ \ref{Subsec:weak}]. From this aspect, our optimization method tends to be more accurate, since it avoids the potential inaccuracy in the damping rates and operators.

\begin{figure}[h!]
    \centering
\includegraphics[width=8.5cm]{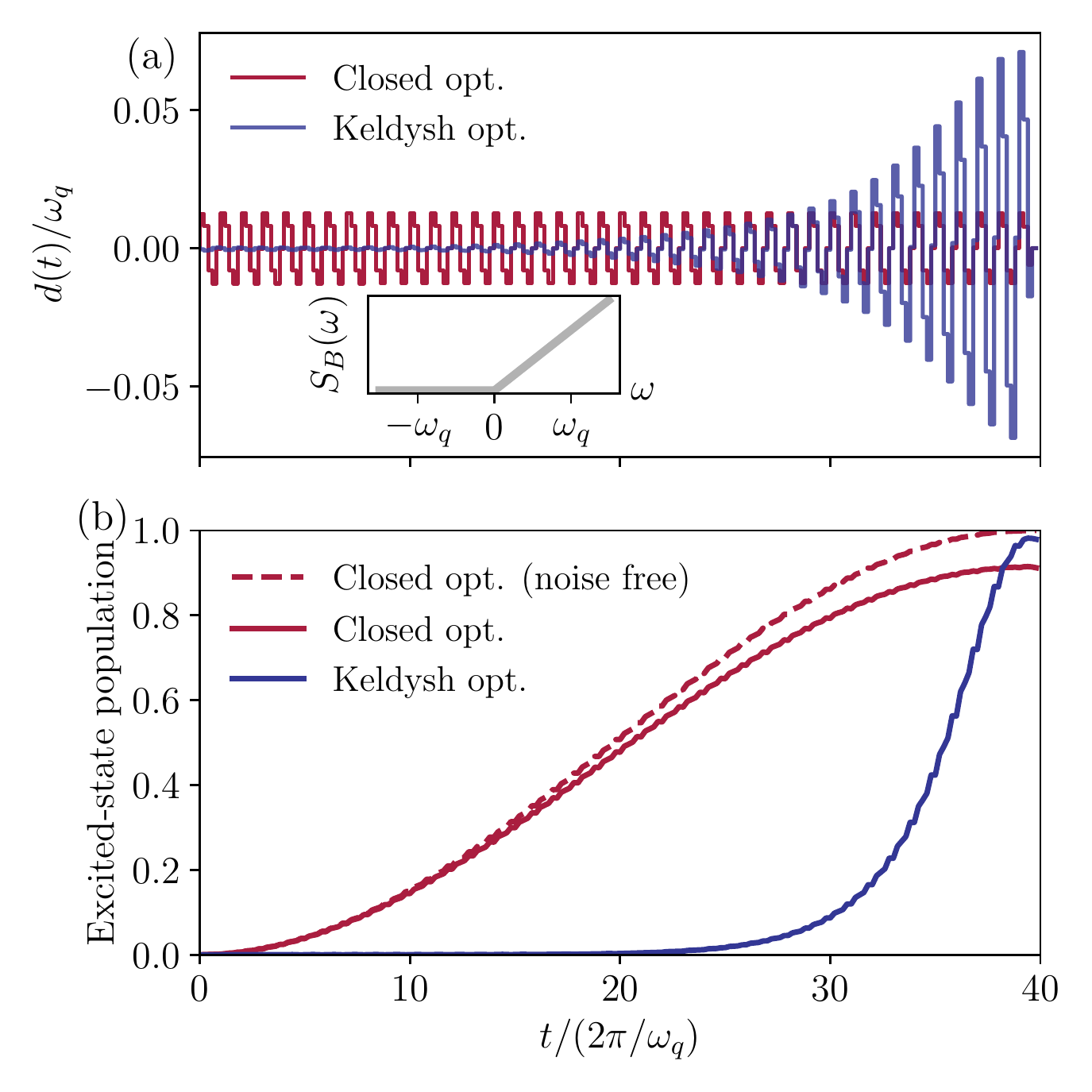}
    \caption{Optimization of the state-transfer fidelity for a qubit coupled to an Ohmic noise bath. (a) shows pulses from both the closed-system  (red) and open-system (blue) optimizations. The spectrum of the Ohmic noise is given in the inset. In (b), we simulate the evolution of the excited-state populations during the $\vert g\rangle \rightarrow \vert e\rangle$ transfer using the two pulses, respectively. The solid curves show the open-system evolution of the population, and the dashed one shows the closed-system evolution. For this simulation, we choose the amplitude of the Ohmic noise as $\mathcal{A}_o = 0.001$.}
    \label{fig:StateTransfer}
\end{figure}

\subsection{Avoiding two-level-system losses in gate operations}
\label{Subsec:gate}

Besides state transfer, the Keldysh-assisted optimizer can also help improve gate fidelities. Especially, if the fidelities of certain intuitive gates are limited by one or several resonance peaks in the noise spectrum, our optimizer can offer solutions that reduce the system sensitivity to noise associated with those peaks. 

To demonstrate this, we consider a noise bath that consists of one Ohmic bath and a few two-level systems (TLSs) \cite{Koch_TLS_FD_thoerem,Muller_tls_review} [see gray curve for the spectrum of the bath in Fig.~\ref{fig:identity} (b)]. These discrete-level defects have been widely believed to limit the coherence times of many solid-state qubits \cite{Muller_tls_review}. Therefore, their mitigation is currently an indispensable task. 

For the concrete simulation, we choose the same Hamiltonian and coupling operator as in the previous case. We consider a situation where the qubit has a fixed frequency but is accidentally in close resonance with the TLSs. For this setup, the gate operations enabled by idling or weakly driving the qubit should suffer significantly from the TLS loss, because the locations of the peaks of $M_k$  overlap with those of the resonance peaks 
(see Fig.~\ref{fig:FilterWeights}).  This situation motivates us to explore pulses that can mitigate the TLS loss.

\begin{figure}[h!]
    \centering
\includegraphics[width=8.5cm]{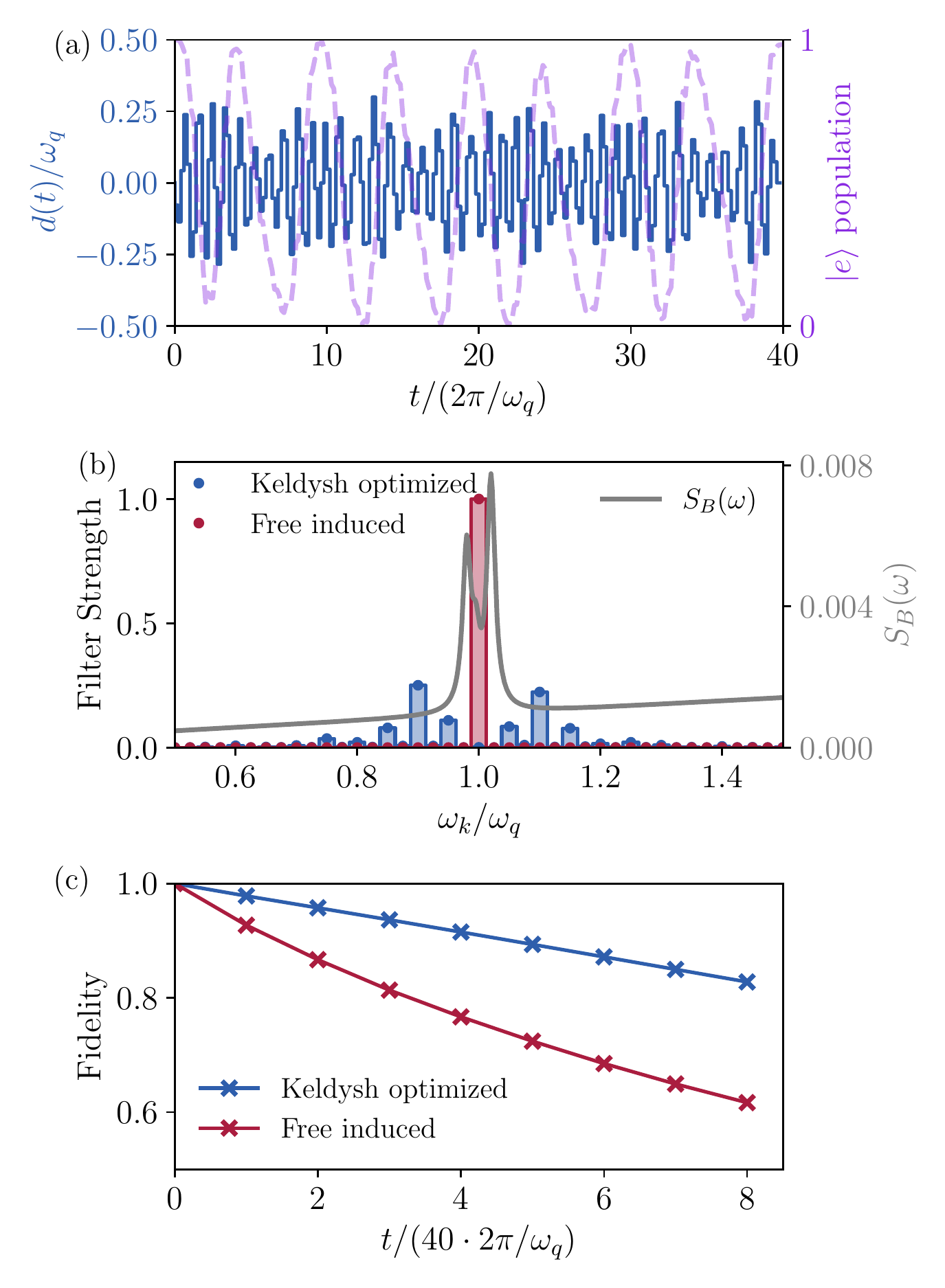}
    \caption{Optimization of the identity gates for a qubit coupled to a few near-resonant TLSs. (a) shows the pulse obtained by the Keldysh-assisted optimal control (blue) and the resulting evolution of the excited-state population (purple) with the qubit initialized in the excited state. (b) plots the filter strengths for the free-induced identity gate (red) and Keldysh-optimized (blue) version. The noise spectrum used for the simulation is also plotted in gray, which is the sum of an Ohmic noise spectrum ($\mathcal{A}_O=3\times 10^{-4}$) and those of three transversely coupled TLSs. (See the form of the spectral function of TLSs in Ref.~\cite{Koch_TLS_FD_thoerem}.) The frequencies of the TLSs are chosen to be close to that of the qubit (0.980, 0.995, and 1.020 $\omega_q$), and their relaxation times are chosen as $ 10\cdot 2\pi/\omega_q$. (c) compares the fidelities of the two identity operations after they are each repeated multiple times.}
    \label{fig:identity}
\end{figure}

Toward this goal, we use the Keldysh-assisted optimal-control technique to optimize the pulse $d(t)$, with the cost function set as the gate infidelity given in Eq.~\eqref{eq:gterror}.  In the following, we focus on optimizing the identity gate as an example, which is the key for the quantum-information storage and multi-qubit operations \cite{Koch_twofluxonium_Gate}. For the simple identity gates enabled by idling the qubit, the interaction between the qubit and the TLSs limits the fidelity of such operation to $E_\mathrm{gate} = 7.0\times 10^{-2}$ over the duration $\tau=40\cdot 2\pi/\omega_q$ [see the overlap of the filter-strength peak of the idle qubit (red) and noise resonance peaks in Fig.~\ref{fig:identity} (b)]. The Keldysh-assisted optimizer, by contrast, proposes to drive the qubit strongly [$d(t)/\omega_q\sim 0.25$] according to the solid curve shown in Fig.~\ref{fig:identity} (a). As a result of the application of this drive, the populations in the two qubit states oscillate over the whole gate period, and return to the original values at the end of the pulse [see dashed curve in (a) for the excited-state population]. These oscillations lead to the appearance of multiple peaks in $M_k$ located away from qubit frequency $\omega_q$ [blue plot in (b)], while the values of $M_k$ at the TLS resonance frequencies are suppressed. As a result, the error in the identity operation is reduced by $3.2\times$ to $E_{\mathrm{gt}} = 2.2\times 10^{-2}$. 

For a clear visual comparison between the two schemes, in (c) we show the identity fidelities after multiple such operations are applied. One can observe that the optimized gate yields much higher fidelity for such repeated applications, which corresponds to a longer effective coherence time in the qubit. We also perform optimization for $X$ and phase gates, and find improvement of a similar magnitude.

\section{Conclusion and outlook}
\label{Sec:conclusion}

In conclusion, we introduce a decoherence model for evaluating errors in a noisy driven system subjected to correlated quantum noise.  The second-order Keldysh expansion and the secular approximation lead to a simple CPTP map \eqref{CPTP} for the system density matrix. Using this map, we study decoherence errors in a variety of quantum systems with both periodic and non-periodic drives. The clear physical picture of the noise sensitivity described by $M_k$ provides useful information for developing noise-mitigation strategies, especially if the noise spectrum is only qualitatively understood but cannot be accurately measured. The simplicity of the map after the secular approximation makes this decoherence model suitable to be integrated with the quantum-optimal-control technique. Using the examples of both state-transfer and single-qubit gate, we show that the combination can help mitigate non-classical and correlated noise in state transfers and gate operations. 

In the future, one may consider using the technique developed in Ref.~\cite{Blais_dyson} for calculating $\hat{U}_s(t)$ for an even more numerically efficient optimization, since that technique is also based on a Dyson series (also the basis for our Keldysh calculation). For capturing higher-order decoherence effects, it is also useful to explore a higher-order CPTP map \cite{Huang_Keldysh}. Finally, our analytically simple map \eqref{CPTP} can provide hints for studying decoherence processes for more complicated systems, e.g., nonlinear oscillators \cite{Kerr-cat_devoret} and composite system in the ultrastrong-coupling regime \cite{Nori_ultra_review}.

\acknowledgements{This material is based upon work supported by the U.S. Department of Energy, Office of Science, National Quantum Information Science Research Centers, Superconducting Quantum Materials and Systems Center (SQMS) under contract number DE-AC02-07CH11359. We thank Peter Groszkowski, Yuxin Wang, Oluwadara Ogunkoya, and \'Angel Rivas for the helpful discussion.}

\appendix
\section{Full second-order expansion and filter functions}\label{ap:filter}

We present more details about the expansion of Eq.~\eqref{Pi2} and the filter functions $I_{k,k'}(\omega)$ in this appendix.

In terms of $\tilde{x}(t)$ and $\tilde{\eta}(t)$, Eq.~\eqref{Pi2} can be expressed as
\begin{align}
    \mathbf{\Sigma}^{(2)}(\tau)\tilde{\rho}_s(0) = &\,(-i)^2\!\int_0^\tau\!\!dt_1\!\int_0^{t_1}\!\!dt_2\,\tilde{x}(t_1)\tilde{x}(t_2)\tilde{\rho}_s(0) \nonumber\\
    &\qquad\quad\times\epsilon^2\mathrm{Tr}_B\{\tilde{\eta}(t_1)\tilde{\eta}(t_2)\tilde{\rho}_B(0)\}\label{eq:(a)}\\
    &+(i)^2\!\int_0^\tau\!\!dt_1\!\int_0^{t_1}\!\!dt_2\,\tilde{\rho}_s(0)\tilde{x}(t_2)\tilde{x}(t_1)\nonumber\\
    &\qquad\quad\times\epsilon^2\mathrm{Tr}_B\{\tilde{\rho}_B(0)\tilde{\eta}(t_2)\tilde{\eta}(t_1)\}\label{eq:(b)}\\
    &+(i)(-i)\!\int_0^\tau\!\!dt_1\!\int_0^{\tau}\!\!dt\,\tilde{x}(t_1)\tilde{\rho}_s(0)\tilde{x}(t_2) \nonumber\\
    &\qquad\quad\times\epsilon^2\mathrm{Tr}_B\{\tilde{\eta}(t_1)\tilde{\rho}_B(0)\tilde{\eta}(t_2)\}.\label{selfenergyfull}
\end{align}
The integrals Eqs.~\eqref{eq:(a)}-\eqref{selfenergyfull} can be conveniently represented by the Keldysh diagrams in Fig.~\ref{fig:Keldysh2time} (see caption for more explanation). These diagrams are useful for collecting integrals in the expansion of $\mathbf{\Sigma}(\tau)$, especially the high-order contributions \cite{Huang_Keldysh}.

\begin{figure}
    \centering
    \includegraphics[width=4.6cm]{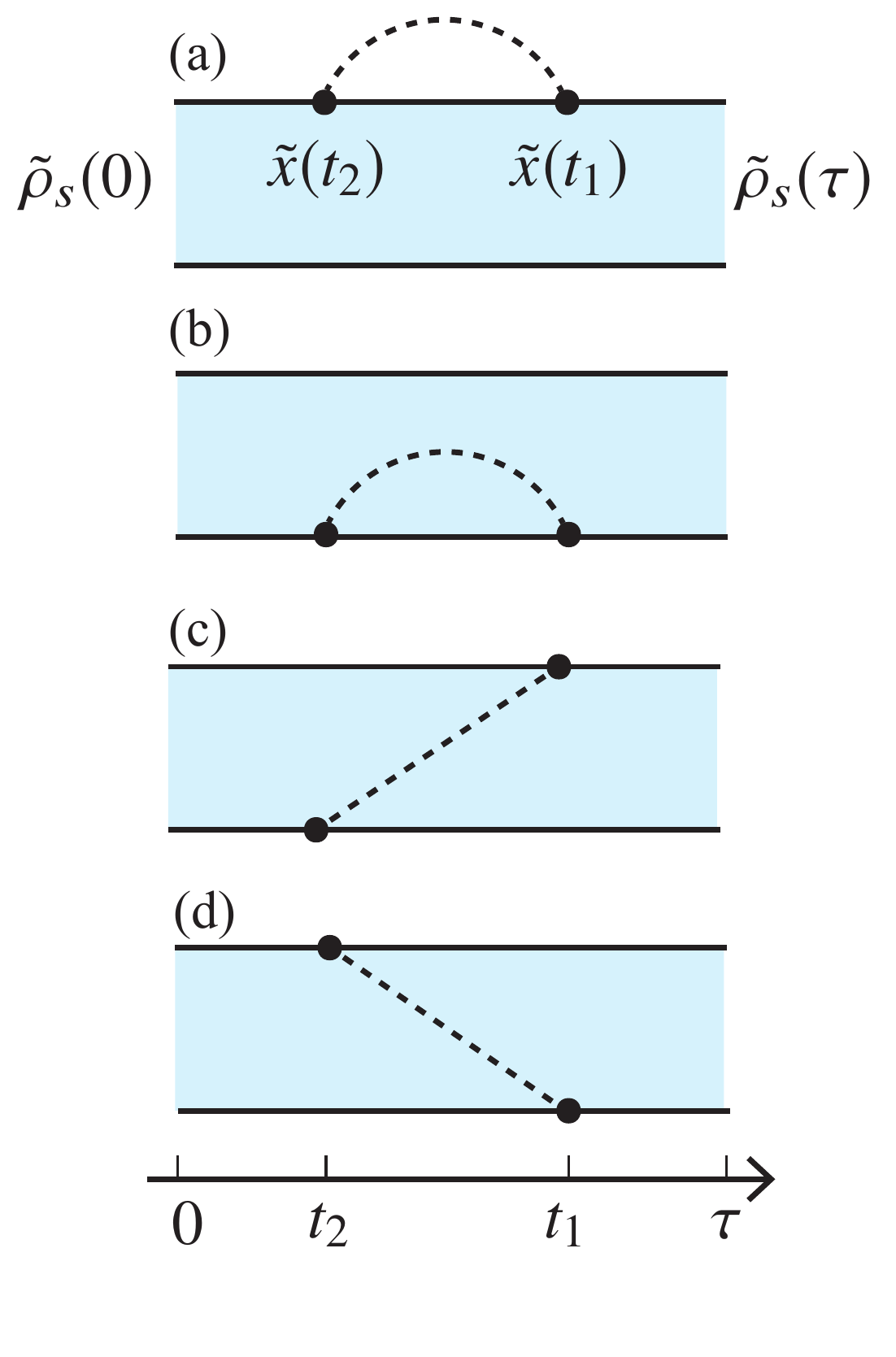}
    \caption{Second-order Keldysh diagrams. They describe how the noisy environment perturbs the evolution of the system density matrix $\tilde{\rho}_s(\tau)$.  (a) and (b) correspond to the integrals Eqs.~\eqref{eq:(a)} and \eqref{eq:(b)}, respectively, while (c) and (d) together represent the integral Eq.~\eqref{selfenergyfull} (note the time ordering in the diagrams). The dashed lines in these diagrams represent the noise correlation functions.}
    \label{fig:Keldysh2time}
\end{figure}

Then, inserting $\tilde{\rho}_B(0) = \hat{\rho}_{B,\mathrm{eq}}$ and the Fourier transformation
\begin{align}
    \epsilon^2\mathrm{Tr}_B\{\hat{\rho}_{B,\mathrm{eq}}\, \tilde{\eta}(t_1)\tilde{\eta}(t_2)\}\!
    =\! \int_{-\infty}^{\infty}\frac{d\omega}{2\pi} S_B(\omega) \exp[-i\omega(t_1-t_2)]\label{correlation_spec}
\end{align}
into Eqs.~\eqref{eq:(a)}-\eqref{selfenergyfull}, we further transform $\mathbf{\Sigma}^{(2)}(\tau)$ into
\begin{align}
     \mathbf{\Sigma}^{(2)}(\tau)\tilde{\rho}_s(0) = &\,-\int_{-\infty}^{\infty}\frac{d\omega}{2\pi}S_B(\omega)\int_0^\tau\!\!dt_1\!\int_0^{t_1}\!\!dt_2\nonumber\\
    &\quad\times \tilde{x}(t_1)\tilde{x}(t_2)\tilde{\rho}_s(0) e^{-i\omega (t_1-t_2)}\nonumber\\
    &\,-\int_{-\infty}^{\infty}\frac{d\omega}{2\pi}S_B(\omega)\int_0^\tau\!\!dt_1\!\int_0^{t_1}\!\!dt_2\nonumber\\
    &\quad\times\tilde{\rho}_s(0)  \tilde{x}(t_2)\tilde{x}(t_1) e^{-i\omega (t_2-t_1)}\nonumber\\
    &\,+\int_{-\infty}^{\infty}\frac{d\omega}{2\pi}S_B(\omega)\int_0^\tau\!\!dt_1\!\int_0^{\tau}\!\!dt_2\nonumber\\
    &\quad\times\tilde{x}(t_1)\tilde{\rho}_s(0)  \tilde{x}(t_2) e^{-i\omega (t_2-t_1)}.\label{selfenergyfull2}
\end{align}

In Sec.\ \ref{Subsec:Fourierexp}, Eq.~\eqref{selfenergyfull2} is further transformed into Eq.~\eqref{map_full}, which is based on the filter function $I_{k,k'}(\omega)$ defined in Eq.~\eqref{eq:ikk'}. Carrying out the double integral in Eq.~\eqref{eq:ikk'}, we find
\begin{align}
    I_{k,k'}(\omega) =&\, \frac{e^{-i\omega\tau}-1}{(\omega-k\omega_p)(k'\omega_p-\omega)}\nonumber\\
    &-\frac{i\tau}{(\omega-k\omega_p)}\delta_{k,k'}. 
    \label{filter_full}
\end{align}
Note that the apparent poles in this expression are all removable. For $k=k'$, Eq.~\eqref{filter_full} is reduced to Eq.~\eqref{IkkKRKI}, which has been studied in detail in Sec.\ \ref{Subsec:Fourierexp}. For off-diagonal ones, we find the inequality
\begin{align}
    |I_{k,k'}(\omega)| < \frac{\tau^2}{2\pi\Big(|k-k'|-\frac{1}{2}\Big)}\label{ampdecay}
\end{align}
by inspecting the first line of Eq.~\eqref{filter_full}. This inequality is referenced in Sec.\ \ref{Subsec:Fourierexp} for justifying the secular approximation.
\section{Connection to Ref.~\cite{Bluhm_filter_functions}}\label{ap:classicalconnection}
We use this appendix to connect our theory to previous filter-function research. Particularly, we show that if only classical noise is present, the map Eq.~\eqref{fullwave} can reproduce some of the formulas used in Ref.~\cite{Bluhm_filter_functions}.

For classical noise, the correlation function $C(t_1,t_2)\equiv \mathrm{Tr}_B\{\hat{\rho}_{B,\mathrm{eq}}\, \tilde{\eta}(t_1)\tilde{\eta}(t_2)\}$ is real-valued, i.e.,
\begin{align}
    C(t_1,t_2) = C^*(t_2,t_1) =  C(t_2,t_1).\label{commute}
\end{align}
This relation implies that the noise spectrum is symmetric, i.e., $S_B(\omega) = S_B(-\omega)$. Then, if we insert Eq.~\eqref{commute} into  Eqs.~\eqref{eq:(a)}-\eqref{selfenergyfull}, we find the self-energy
\begin{align}
    \mathbf{\Sigma}^{(2)}(\tau)\tilde{\rho}_s(0) = &\,-\!\int_0^t\!\!dt_1\!\int_0^{t_1}\!\!dt_2\,\tilde{x}(t_1)\tilde{x}(t_2)\tilde{\rho}_s(0) C(t_1,t_2)\nonumber\\
    &-\!\int_0^t\!\!dt_1\!\int_0^{t_1}\!\!dt_2\,\tilde{\rho}_s(0)\tilde{x}(t_2)\tilde{x}(t_1) C(t_1,t_2)\nonumber\\
    &+\!\int_0^t\!\!dt_1\!\int_0^{t}\!\!dt_2\,\tilde{x}(t_1)\tilde{\rho}_s(0)\tilde{x}(t_2) C(t_1,t_2).
\end{align}
Inserting this quantity into the approximated map
 $\mathbf{\Pi}(\tau)\approx\mathbf{\Pi}^{(0)}(\tau) + \mathbf{\Sigma}^{(2)}(\tau)$, we recover the \textit{noise-averaged quantum process} in Ref.~\cite{Bluhm_filter_functions}.

\section{Bath correlation time and spectral variation}\label{ap:correlation}

In this appendix, we investigate the relation between the variation of $S_B(\omega)$ and the bath correlation time. Such a relation is useful for interpreting condition \ding{173} as a comparison between the bath correlation time and the system evolution time.

The variation of $S_B(\omega)$ can be roughly quantified by the second-order derivative of the spectrum. Specifically, we define the spectral roughness $R(\omega)$ by
\begin{align}
    R(\omega) \equiv \frac{|d^2S_B(\omega)/d\omega^2|}{|S_B(\omega)|}.
\end{align}
To relate this quantity to the correlation time, we insert the inverse Fourier transform of $S_B(\omega)$ into the expression $R(\omega)$ and express it as
\begin{align}
    R(\omega) =&\, \frac{\Big|\int_{-\infty}^{\infty}dt \,C(t,0)t^2\,e^{i\omega t}\Big|}{\big|\int_{-\infty}^{\infty}dt \,C(t,0) e^{i\omega t}\big|}.\label{eq:rwrrw}
\end{align}
The right-hand side of Eq.~\eqref{eq:rwrrw} appears to be related to the bath correlation time. To understand this expression more clearly, we consider a two-level-system defect as an example. The time-domain correlation function of such a two-level system (at zero temperature) has the form \cite{Koch_TLS_FD_thoerem} 
\begin{align}
    C(t,0) = C(0,0) e^{-i\omega_t t -  t/T_t},
\end{align}
where $\omega_t$ and $T_t$ are its resonance frequency and coherence time (usually considered as its correlation time), respectively. For this function, the right-hand side of Eq.~\eqref{eq:rwrrw} is evaluated as
\begin{align}
    \frac{\Big|\int_{-\infty}^{\infty}dt \,C(t,0)t^2\,e^{i\omega t}\Big|}{\big|\int_{-\infty}^{\infty}dt \,C(t,0) e^{i\omega t}\big|} = {\frac{{\Big| 2/T_t^2 - 6(\omega-\omega_t)^2  \Big|}}{\big[1/T_t^2 + (\omega-\omega_t)^2 \big]^2}},
\end{align}
which equals $2T_t^2$ for $\omega=\omega_t$. 

Therefore, it is reasonable to define a frequency-dependent bath correlation time
\begin{align}
    \tau_B(\omega) =\sqrt{ \frac{\Big|\int_{-\infty}^{\infty}dt \,C(t,0)t^2\,e^{i\omega t}\Big|}{2\big|\int_{-\infty}^{\infty}dt \,C(t,0) e^{i\omega t}\big|}}.
\end{align}
Such a definition leads to the relation
\begin{align}
    R(\omega) = 2\tau^2_B(\omega),
\end{align}
which indicates that more significant variations in $S_B(\omega)$ correspond to longer bath correlation times. 

\section{Floquet master equation for a weakly driven qubit}\label{floquet_weak}

In this appendix, we use the Markovian Floquet master equation to explain the appearance of the side peaks in Fig.\ \ref{fig:FilterWeights} (b).

The two Floquet states for the weakly driven qubit are 
\begin{align}
    \vert w_{\pm}(t)\rangle = \frac{1}{\sqrt{2}}\Big[\vert g\rangle  \pm\vert e\rangle e^{-i{\omega_qt}} \Big],\nonumber
\end{align}
and their quasi-energies are $\varepsilon_{\pm}=\pm d/2-\omega_q/2$. Inserting them into Eq.~\eqref{eq:floquet_xtilde}, we can again find the expression of the rotated coupling operator Eq.~\eqref{xtdrivenres}. 

Following the derivation of the Floquet master equation, we extract the transition frequencies $\omega_L$ and their corresponding damping operators $\tilde{x}(\omega_L)$ from Eq.~\eqref{xtdrivenres}. Then, the Lindbladian for the Markovian Floquet master equation is given by 
\begin{align}
    \mathcal{L} = &\,  S_B(\omega_q) \mathbb{D}\Big[\frac{\hat{\sigma}_x}{2}\Big] +\sum_{\pm}  S_B(\omega_q\pm d) \mathbb{D}\Big[\frac{\mp\hat{\sigma}_z- i\hat{\sigma}_y}{4}\Big]\nonumber\\
     &+S_B(-\omega_q) \mathbb{D}\Big[\frac{\hat{\sigma}_x}{2}\Big] +\sum_{\pm}  S_B(-\omega_q\mp d) \mathbb{D}\Big[\frac{\mp\hat{\sigma}_z+ i\hat{\sigma}_y}{4}\Big]\nonumber\\
     &+\text{Lamb-shift terms}.
\end{align}
The damping terms present in this map indeed capture the noise channels predicted in Fig.~\ref{fig:FilterWeights} (b).

\section{Driven harmonic oscillator}\label{ap:oscillator}

In the main text, the examples we present are limited to qubits with only two levels. Here, we demonstrate the applicability of our framework in a quantum harmonic oscillator, which has infinite levels.

The system Hamiltonian of this oscillator is specified by
\begin{align}
    \hat{H}_s(t) = \omega_r \hat{a}^\dagger \hat{a} + d(t)(\hat{a}^\dagger+\hat{a}),
\end{align}
and the coupling operator for the oscillator is $\hat{x}=\hat{a}+\hat{a}^\dagger$. For this linear system, the closed-system propagator can be analytically derived as \cite{Blais_cqed_review}
\begin{align}
    \hat{U}_s(t)  = e^{\alpha(t)\hat{a}^\dagger -\alpha^*(t)\hat{a}}e^{-i\omega_r\hat{a}^\dagger\hat{a}t}e^{-i\Phi(t)},
\end{align}
where the displacement $\alpha$ is calculated by
\begin{align}
    i\dot{\alpha}(t) = \omega_r\alpha(t) + d(t)\label{alphat},
\end{align}
and the additional phase acquired is given by
\begin{align}
    \Phi(t) = -\int_0^t dt'\Bigg[\omega_r|\alpha|^2 +\frac{1}{2}i(\alpha\dot{\alpha}^* - \alpha^*\dot{\alpha})\Bigg].
\end{align}

 In the interaction picture, the coupling operator is transformed as
\begin{align}
    \tilde{x}(t) = [\hat{a}e^{-i\omega_rt} +\alpha(t)]+\mathrm{H.c.},
\end{align}
which leads to only two frequency components, $\tilde{x}(\omega_r) = \hat{a}+{\alpha}$ and $\tilde{x}(-\omega_r) = {a}^\dagger+\bar{\alpha}^*$. We note that the c-numbers $\alpha(t),\alpha^*(t)$  in $\tilde{x}(\pm\omega_r)$ only contribute to the Lamb-shift terms to the map \eqref{CPTP}, which we choose to omit due to the weak noise strength.

Then, if we again assume that the two conditions \ding{172} and \ding{173} in Sec.\ \ref{Subsec:static} hold, the self-energy is given by
\begin{align}
    \mathbf{\Sigma}^{(2)    }_{\mathrm{CP}}(\tau) = \tau\big\{S_B(\omega_r)\mathbb{D}[\hat{a}] + S_B(-\omega_r)\mathbb{D}[\hat{a}^\dagger]\big\},
\end{align}
identical to the prediction by the Lindblad master equation. The message sent by such analysis is that, the Lindblad map is a good approximation for the harmonic oscillator under arbitrary linear drives, as long as $\tau_B$ and $\tau_S\sim2\pi/\omega_r$ are much smaller compared to $\tau\sim\tau_R$. Such a conclusion is in clear contrast to those in Sec.\ \ref{Subsec:weak} and \ref{Subsec:floquet} for the driven qubits. Note that this conclusion may be invalid if nonlinearity in the cavity is induced by its coupling to qubits \cite{Kerr-cat_devoret}. This conclusion may also be invalid if the drive also affects the noise bath \cite{Kirchmair_nonMarkovian}, which results in a varying $S_B(\omega)$ during the drive time.


\begin{thebibliography}{10}

\bibitem{Preskill_NISQ_review}
John Preskill.
\newblock ``\textit{Quantum Computing in the NISQ Era and Beyond}''.
\newblock \href{https://dx.doi.org/10.22331/q-2018-08-06-79}{Quantum {\bf 2},
  79}~(2018).

\bibitem{Lidar_coarse_grain}
Christian Majenz, Tameem Albash, Heinz-Peter Breuer, and Daniel~A. Lidar.
\newblock ``\textit{Coarse Graining Can Beat the Rotating-Wave Approximation in
  Quantum Markovian Master Equations}''.
\newblock \href{https://dx.doi.org/10.1103/PhysRevA.88.012103}{Phys. Rev. A
  {\bf 88}, 012103}~(2013).

\bibitem{Lidar_coarse_grain_drive}
Evgeny Mozgunov and Daniel Lidar.
\newblock ``\textit{Completely Positive Master Equation for Arbitrary Driving
  and Small Level Spacing}''.
\newblock \href{https://dx.doi.org/10.22331/q-2020-02-06-227}{{Quantum} {\bf
  4}, 227}~(2020).

\bibitem{Groszkowski_Magnus_master}
Peter Groszkowski, Alireza Seif, Jens Koch, and A.~A. Clerk.
\newblock ``\textit{Simple Master Equations for Describing Driven Systems
  Subject to Classical Non-{M}arkovian Noise}''.
\newblock \href{https://dx.doi.org/10.22331/q-2023-04-06-972}{{Quantum} {\bf
  7}, 972}~(2023).

\bibitem{Dynamical_sweet_spot}
Ziwen Huang, Pranav~S. Mundada, Andr\'as Gyenis, David~I. Schuster, Andrew~A.
  Houck, and Jens Koch.
\newblock ``\textit{Engineering Dynamical Sweet Spots to Protect Qubits from
  $1/f$ Noise}''.
\newblock \href{https://dx.doi.org/10.1103/PhysRevApplied.15.034065}{Phys. Rev.
  Appl. {\bf 15}, 034065}~(2021).

\bibitem{Green_filer_func}
Todd~J. Green, Jarrah Sastrawan, Hermann Uys, and Michael~J. Biercuk.
\newblock ``\textit{Arbitrary Quantum Control of Qubits in the Presence of
  Universal Noise}''.
\newblock \href{https://dx.doi.org/10.1088/1367-2630/15/9/095004}{New J. Phys.
  {\bf 15}, 095004}~(2013).

\bibitem{Norris_filter_func}
Vivian Maloney, Yasuo Oda, Gregory Quiroz, B.~David Clader, and Leigh~M.
  Norris.
\newblock ``\textit{Qubit Control Noise Spectroscopy with Optimal Suppression
  of Dephasing}''.
\newblock \href{https://dx.doi.org/10.1103/PhysRevA.106.022425}{Phys. Rev. A
  {\bf 106}, 022425}~(2022).

\bibitem{Didier_dynamical_sweet_spot}
Nicolas Didier.
\newblock ``\textit{Flux Control of Superconducting Qubits at Dynamical Sweet
  Spot}''~(2019).
\newblock \href{https://doi.org/10.48550/arXiv.1912.09416}{arXiv.1912.09416}.

\bibitem{Muller_Keldysh}
Clemens M\"uller and Thomas~M. Stace.
\newblock ``\textit{Deriving Lindblad Master Equations with Keldysh Diagrams:
  Correlated Gain and Loss in Higher Order Perturbation Theory}''.
\newblock \href{https://dx.doi.org/10.1103/PhysRevA.95.013847}{Phys. Rev. A
  {\bf 95}, 013847}~(2017).

\bibitem{Fogedby_field_theory_lindblad}
Hans~C. Fogedby.
\newblock ``\textit{Field-Theoretical Approach to Open Quantum Systems and the
  Lindblad Equation}''.
\newblock \href{https://dx.doi.org/10.1103/PhysRevA.106.022205}{Phys. Rev. A
  {\bf 106}, 022205}~(2022).

\bibitem{TrushechkinUnified}
Anton Trushechkin.
\newblock ``\textit{Unified Gorini-Kossakowski-Lindblad-Sudarshan Quantum
  Master Equation Beyond the Secular Approximation}''.
\newblock \href{https://dx.doi.org/10.1103/PhysRevA.103.062226}{Phys. Rev. A
  {\bf 103}, 062226}~(2021).

\bibitem{Vcchini_lindblad_generalized}
Bassano Vacchini.
\newblock ``\textit{Generalized Master Equations Leading to Completely Positive
  Dynamics}''.
\newblock \href{https://dx.doi.org/10.1103/PhysRevLett.117.230401}{Phys. Rev.
  Lett. {\bf 117}, 230401}~(2016).

\bibitem{Rivas_coarse_graining}
\'Angel Rivas.
\newblock ``\textit{Refined Weak-Coupling Limit: Coherence, Entanglement, and
  Non-Markovianity}''.
\newblock \href{https://dx.doi.org/10.1103/PhysRevA.95.042104}{Phys. Rev. A
  {\bf 95}, 042104}~(2017).

\bibitem{Alicki_coarse_graining}
Robert Alicki.
\newblock ``\textit{Master Equations for a Damped Nonlinear Oscillator and the
  Validity of the Markovian Approximation}''.
\newblock \href{https://dx.doi.org/10.1103/PhysRevA.40.4077}{Phys. Rev. A {\bf
  40}, 4077--4081}~(1989).

\bibitem{Brandes_coarse_graining}
Gernot Schaller and Tobias Brandes.
\newblock ``\textit{Preservation of Positivity by Dynamical Coarse Graining}''.
\newblock \href{https://dx.doi.org/10.1103/PhysRevA.78.022106}{Phys. Rev. A
  {\bf 78}, 022106}~(2008).

\bibitem{Green_filter_func_high_order}
Todd Green, Hermann Uys, and Michael~J. Biercuk.
\newblock ``\textit{High-Order Noise Filtering in Nontrivial Quantum Logic
  Gates}''.
\newblock \href{https://dx.doi.org/10.1103/PhysRevLett.109.020501}{Phys. Rev.
  Lett. {\bf 109}, 020501}~(2012).

\bibitem{Bluhm_filter_functions}
Pascal Cerfontaine, Tobias Hangleiter, and Hendrik Bluhm.
\newblock ``\textit{Filter Functions for Quantum Processes under Correlated
  Noise}''.
\newblock \href{https://dx.doi.org/10.1103/PhysRevLett.127.170403}{Phys. Rev.
  Lett. {\bf 127}, 170403}~(2021).

\bibitem{Bluhm_filter_functions_PRR}
Tobias Hangleiter, Pascal Cerfontaine, and Hendrik Bluhm.
\newblock ``\textit{Filter-function Formalism and Software Package to Compute
  Quantum Processes of Gate Sequences for Classical Non-Markovian Noise}''.
\newblock \href{https://dx.doi.org/10.1103/PhysRevResearch.3.043047}{Phys. Rev.
  Res. {\bf 3}, 043047}~(2021).

\bibitem{Oliver_flux_qubit_dd}
Jonas Bylander, Simon Gustavsson, Fei Yan, Fumiki Yoshihara, Khalil Harrabi,
  George Fitch, David~G. Cory, Yasunobu Nakamura, Jaw-Shen Tsai, and William~D.
  Oliver.
\newblock ``\textit{Noise Spectroscopy Through Dynamical Decoupling with a
  Superconducting Flux Qubit}''.
\newblock \href{https://dx.doi.org/10.1038/nphys1994}{Nat. Phys. {\bf 7},
  565--570}~(2011).

\bibitem{Viola_noise_spectroscopy}
Gerardo~A. Paz-Silva, Leigh~M. Norris, and Lorenza Viola.
\newblock ``\textit{Multiqubit Spectroscopy of Gaussian Quantum Noise}''.
\newblock \href{https://dx.doi.org/10.1103/PhysRevA.95.022121}{Phys. Rev. A
  {\bf 95}, 022121}~(2017).

\bibitem{Zeng_filter_func}
Kevin Schultz, Ryan LaRose, Andrea Mari, Gregory Quiroz, Nathan Shammah,
  B.~David Clader, and William~J. Zeng.
\newblock ``\textit{Impact of Time-Correlated Noise on Zero-Noise
  Extrapolation}''.
\newblock \href{https://dx.doi.org/10.1103/PhysRevA.106.052406}{Phys. Rev. A
  {\bf 106}, 052406}~(2022).

\bibitem{Shnirman_Keldysh_dephasing}
Yuriy Makhlin and Alexander Shnirman.
\newblock ``\textit{Dephasing of Solid-State Qubits at Optimal Points}''.
\newblock \href{https://dx.doi.org/10.1103/PhysRevLett.92.178301}{Phys. Rev.
  Lett. {\bf 92}, 178301}~(2004).

\bibitem{Leung_optimal_control}
Nelson Leung, Mohamed Abdelhafez, Jens Koch, and David Schuster.
\newblock ``\textit{Speedup for Quantum Optimal Control from Automatic
  Differentiation Based on Graphics Processing Units}''.
\newblock \href{https://dx.doi.org/10.1103/PhysRevA.95.042318}{Phys. Rev. A
  {\bf 95}, 042318}~(2017).

\bibitem{Dynamical_sweet_spot_exp}
Pranav~S. Mundada, Andr\'as Gyenis, Ziwen Huang, Jens Koch, and Andrew~A.
  Houck.
\newblock ``\textit{Floquet-Engineered Enhancement of Coherence Times in a
  Driven Fluxonium Qubit}''.
\newblock \href{https://dx.doi.org/10.1103/PhysRevApplied.14.054033}{Phys. Rev.
  Appl. {\bf 14}, 054033}~(2020).

\bibitem{Didier_dynamical_sweet_spot_exp}
Joseph~A. Valery, Shoumik Chowdhury, Glenn Jones, and Nicolas Didier.
\newblock ``\textit{Dynamical Sweet Spot Engineering via Two-Tone Flux
  Modulation of Superconducting Qubits}''.
\newblock \href{https://dx.doi.org/10.1103/PRXQuantum.3.020337}{PRX Quantum
  {\bf 3}, 020337}~(2022).

\bibitem{Clerk_non_Gaussian}
Yu-Xin Wang and A.~A. Clerk.
\newblock ``\textit{Spectral Characterization of Non-Gaussian Quantum Noise:
  Keldysh Approach and Application to Photon Shot Noise}''.
\newblock \href{https://dx.doi.org/10.1103/PhysRevResearch.2.033196}{Phys. Rev.
  Res. {\bf 2}, 033196}~(2020).

\bibitem{Huang_Keldysh}
Ziwen Huang, Xinyuan You, Ugur Alyanak, Alexander Romanenko, Anna Grassellino,
  and Shaojiang Zhu.
\newblock ``\textit{High-Order Qubit Dephasing at Sweet Spots by Non-Gaussian
  Fluctuators: Symmetry Breaking and Floquet Protection}''.
\newblock \href{https://dx.doi.org/10.1103/PhysRevApplied.18.L061001}{Phys.
  Rev. Appl. {\bf 18}, L061001}~(2022).

\bibitem{Lindblad}
G.~Lindblad.
\newblock ``\textit{On the Generators of Quantum Dynamical Semigroups}''.
\newblock \href{https://dx.doi.org/10.1007/BF01608499}{Commun. Math. Phys. {\bf
  48}, 119--130}~(1976).

\bibitem{Breuer_open_quantum_system}
Heinz-Peter Breuer and Francesco Petruccione.
\newblock ``\textit{The Theory of Open Quantum Systems}''.
\newblock
  \href{https://dx.doi.org/10.1093/acprof:oso/9780199213900.001.0001}{Chapter~3,
  pages 125--131}.
\newblock Oxford University Press, New York. ~(2007).

\bibitem{abdelhafez2019quantum}
Mohamed~Ragab Abdelhafez.
\newblock ``\textit{Quantum Optimal Control Using Automatic Differentiation}''.
\newblock \href{https://dx.doi.org/10.6082/uchicago.2028}{PhD thesis}.
\newblock The University of Chicago.
\newblock ~(2019).

\bibitem{BLANESMagnus}
S.~Blanes, F.~Casas, J.~A. Oteo, and J.~Ros.
\newblock ``\textit{The Magnus Expansion and Some of Its Applications}''.
\newblock
  \href{https://dx.doi.org/https://doi.org/10.1016/j.physrep.2008.11.001}{Phys.
  Rep. {\bf 470}, 151--238}~(2009).

\bibitem{qutip}
J.~R. Johansson, P.~D. Nation, and Franco Nori.
\newblock ``\textit{QuTiP 2: A Python Framework for the Dynamics of Open
  Quantum Systems}''.
\newblock \href{https://dx.doi.org/10.1016/j.cpc.2012.11.019}{Comp. Phys. Comm.
  {\bf 184}, 1234}~(2013).

\bibitem{Oliver_rotating_frame_relaxation}
Fei Yan, Simon Gustavsson, Jonas Bylander, Xiaoyue Jin, Fumiki Yoshihara,
  David~G. Cory, Yasunobu Nakamura, Terry~P. Orlando, and William~D. Oliver.
\newblock ``\textit{Rotating-Frame Relaxation as a Noise Spectrum Analyser of a
  Superconducting Qubit Undergoing Driven Evolution}''.
\newblock \href{https://dx.doi.org/10.1038/ncomms3337}{Nat. Commun. {\bf 4},
  2337}~(2013).

\bibitem{Ithier_decoherence_analysis}
G.~Ithier, E.~Collin, P.~Joyez, P.~J. Meeson, D.~Vion, D.~Esteve, F.~Chiarello,
  A.~Shnirman, Y.~Makhlin, J.~Schriefl, and G.~Sch\"on.
\newblock ``\textit{Decoherence in a Superconducting Quantum Bit Circuit}''.
\newblock \href{https://dx.doi.org/10.1103/PhysRevB.72.134519}{Phys. Rev. B
  {\bf 72}, 134519}~(2005).

\bibitem{Fluxonium_hc}
Long~B. Nguyen, Yen-Hsiang Lin, Aaron Somoroff, Raymond Mencia, Nicholas
  Grabon, and Vladimir~E. Manucharyan.
\newblock ``\textit{High-Coherence Fluxonium Qubit}''.
\newblock \href{https://dx.doi.org/10.1103/PhysRevX.9.041041}{Phys. Rev. X {\bf
  9}, 041041}~(2019).

\bibitem{Groszkowski_Zero_pi_theory}
Peter Groszkowski, A.~Di~Paolo, A.~L. Grimsmo, A.~Blais, D.~I. Schuster, A.~A.
  Houck, and Jens Koch.
\newblock ``\textit{Coherence Properties of the 0-$\pi$ Qubit}''.
\newblock \href{https://dx.doi.org/10.1088/1367-2630/aab7cd}{New J. Phys. {\bf
  20}, 043053}~(2018).

\bibitem{Rodriguez-Rosario_CPTP}
C\'esar~A. Rodr\'iguez-Rosario, Kavan Modi, Aik meng Kuah, Anil Shaji, and
  E.~C.~G. Sudarshan.
\newblock ``\textit{Completely Positive Maps and Classical Correlations}''.
\newblock \href{https://dx.doi.org/10.1088/1751-8113/41/20/205301}{J. Phys. A:
  Math. Theor. {\bf 41}, 205301}~(2008).

\bibitem{Blais_Floquet}
Anthony Gandon, Camille Le~Calonnec, Ross Shillito, Alexandru Petrescu, and
  Alexandre Blais.
\newblock ``\textit{Engineering, Control, and Longitudinal Readout of Floquet
  Qubits}''.
\newblock \href{https://dx.doi.org/10.1103/PhysRevApplied.17.064006}{Phys. Rev.
  Appl. {\bf 17}, 064006}~(2022).

\bibitem{Siddiqi_floquet_qubit}
Long~B. Nguyen, Yosep Kim, Akel Hashim, Noah Goss, Brian Marinelli, Bibek
  Bhandari, Debmalya Das, Ravi~K. Naik, John~Mark Kreikebaum, Andrew~N. Jordan,
  David~I. Santiago, and Irfan Siddiqi.
\newblock ``\textit{Programmable Heisenberg Interactions Between Floquet
  Qubits}''~(2022).
\newblock \href{https://doi.org/10.48550/arXiv.2211.10383}{arXiv:2211.10383}.

\bibitem{Trif_spin_qubit_floquet}
Sarath Prem, Marcin~M. Wysoki\'nski, and Mircea Trif.
\newblock ``\textit{Longitudinal Coupling Between Electrically Driven
  Spin-Qubits and a Resonator}''~(2023).
\newblock \href{https://doi.org/10.48550/arXiv.2301.10163}{arXiv:2301.10163}.

\bibitem{Lupascu_flux_qubit_Floquet}
Chunqing Deng, Jean-Luc Orgiazzi, Feiruo Shen, Sahel Ashhab, and Adrian
  Lupascu.
\newblock ``\textit{Observation of Floquet States in a Strongly Driven
  Artificial Atom}''.
\newblock \href{https://dx.doi.org/10.1103/PhysRevLett.115.133601}{Phys. Rev.
  Lett. {\bf 115}, 133601}~(2015).

\bibitem{koch2022quantum}
Christiane~P Koch, Ugo Boscain, Tommaso Calarco, Gunther Dirr, Stefan Filipp,
  Steffen~J Glaser, Ronnie Kosloff, Simone Montangero, Thomas
  Schulte-Herbr{\"u}ggen, Dominique Sugny, et~al.
\newblock ``\textit{Quantum Optimal Control in Quantum Technologies. Strategic
  Report on Current Status, Visions and Goals for Research in Europe}''.
\newblock
  \href{https://dx.doi.org/https://doi.org/10.1140/epjqt/s40507-022-00138-x}{EPJ
  Quantum Technol. {\bf 9}, 19}~(2022).

\bibitem{gunther2021quandary}
Stefanie G{\"u}nther, N.~Anders Petersson, and Jonathan~L. DuBois.
\newblock ``\textit{Quandary: An Open-Source C++ Package for High-Performance
  Optimal Control of Open Quantum Systems}''.
\newblock In 2021 IEEE/ACM Second International Workshop on Quantum Computing
  Software (QCS).
\newblock \href{https://dx.doi.org/10.1109/QCS54837.2021.00014}{Pages 88--98}.
\newblock Los Alamitos, CA, USA~(2021). IEEE Computer Society.

\bibitem{abdelhafez2019gradient}
Mohamed Abdelhafez, David~I. Schuster, and Jens Koch.
\newblock ``\textit{Gradient-Based Optimal Control of Open Quantum Systems
  Using Quantum Trajectories and Automatic Differentiation}''.
\newblock \href{https://dx.doi.org/10.1103/PhysRevA.99.052327}{Phys. Rev. A
  {\bf 99}, 052327}~(2019).

\bibitem{Schuster_robust_optimal_control}
Thomas Propson, Brian~E. Jackson, Jens Koch, Zachary Manchester, and David~I.
  Schuster.
\newblock ``\textit{Robust Quantum Optimal Control with Trajectory
  Optimization}''.
\newblock \href{https://dx.doi.org/10.1103/PhysRevApplied.17.014036}{Phys. Rev.
  Appl. {\bf 17}, 014036}~(2022).

\bibitem{Biercuk_optimal_control}
A.~Soare, H.~Ball, D.~Hayes, J.~Sastrawan, M.~C. Jarratt, J.~J. McLoughlin,
  X.~Zhen, T.~J. Green, and M.~J. Biercuk.
\newblock ``\textit{Experimental Noise Filtering by Quantum Control}''.
\newblock \href{https://dx.doi.org/10.1038/nphys3115}{Nat. Phys. {\bf 10},
  825--829}~(2014).

\bibitem{qcontrol_filter_func}
Harrison Ball, Michael~J Biercuk, Andre R~R Carvalho, Jiayin Chen, Michael
  Hush, Leonardo A~De Castro, Li~Li, Per~J Liebermann, Harry~J Slatyer, Claire
  Edmunds, Virginia Frey, Cornelius Hempel, and Alistair Milne.
\newblock ``\textit{Software Tools for Quantum Control: Improving Quantum
  Computer Performance Through Noise and Error Suppression}''.
\newblock \href{https://dx.doi.org/10.1088/2058-9565/abdca6}{Quantum Sci.
  Technol. {\bf 6}, 044011}~(2021).

\bibitem{Du_optctrl_cnot}
Tianyu Xie, Zhiyuan Zhao, Shaoyi Xu, Xi~Kong, Zhiping Yang, Mengqi Wang,
  Ya~Wang, Fazhan Shi, and Jiangfeng Du.
\newblock ``\textit{99.92\%-Fidelity CNOT Gates in Solids by Noise
  Filtering}''.
\newblock \href{https://dx.doi.org/10.1103/PhysRevLett.130.030601}{Phys. Rev.
  Lett. {\bf 130}, 030601}~(2023).

\bibitem{Bluhm_optimal_control}
Isabel Nha~Minh Le, Julian~D. Teske, Tobias Hangleiter, Pascal Cerfontaine, and
  Hendrik Bluhm.
\newblock ``\textit{Analytic Filter-Function Derivatives for Quantum Optimal
  Control}''.
\newblock \href{https://dx.doi.org/10.1103/PhysRevApplied.17.024006}{Phys. Rev.
  Appl. {\bf 17}, 024006}~(2022).

\bibitem{khaneja2005optimal}
Navin Khaneja, Timo Reiss, Cindie Kehlet, Thomas Schulte-Herbr{\"u}ggen, and
  Steffen~J Glaser.
\newblock ``\textit{Optimal Control of Coupled Spin Dynamics: Design of NMR
  Pulse Sequences by Gradient Ascent Algorithms}''.
\newblock
  \href{https://dx.doi.org/https://doi.org/10.1016/j.jmr.2004.11.004}{J. Magn.
  Reson. {\bf 172}, 296--305}~(2005).

\bibitem{Koch_TLS_FD_thoerem}
Xinyuan You, Aashish~A. Clerk, and Jens Koch.
\newblock ``\textit{Positive- and Negative-Frequency Noise from an Ensemble of
  Two-Level Fluctuators}''.
\newblock \href{https://dx.doi.org/10.1103/PhysRevResearch.3.013045}{Phys. Rev.
  Res. {\bf 3}, 013045}~(2021).

\bibitem{Muller_tls_review}
Clemens Müller, Jared~H Cole, and Jürgen Lisenfeld.
\newblock ``\textit{Towards Understanding Two-Level-Systems in Amorphous
  Solids: Insights from Quantum Circuits}''.
\newblock \href{https://dx.doi.org/10.1088/1361-6633/ab3a7e}{Rep. Prog. Phys.
  {\bf 82}, 124501}~(2019).

\bibitem{Koch_twofluxonium_Gate}
D.~K. Weiss, Helin Zhang, Chunyang Ding, Yuwei Ma, David~I. Schuster, and Jens
  Koch.
\newblock ``\textit{Fast High-Fidelity Gates for Galvanically-Coupled Fluxonium
  Qubits Using Strong Flux Modulation}''.
\newblock \href{https://dx.doi.org/10.1103/PRXQuantum.3.040336}{PRX Quantum
  {\bf 3}, 040336}~(2022).

\bibitem{Blais_dyson}
Ross Shillito, Jonathan~A. Gross, Agustin Di~Paolo, \'Elie Genois, and
  Alexandre Blais.
\newblock ``\textit{Fast and Differentiable Simulation of Driven Quantum
  Systems}''.
\newblock \href{https://dx.doi.org/10.1103/PhysRevResearch.3.033266}{Phys. Rev.
  Res. {\bf 3}, 033266}~(2021).

\bibitem{Kerr-cat_devoret}
A.~Grimm, N.~E. Frattini, S.~Puri, S.~O. Mundhada, S.~Touzard, M.~Mirrahimi,
  S.~M. Girvin, S.~Shankar, and M.~H. Devoret.
\newblock ``\textit{Stabilization and Operation of a Kerr-Cat Qubit}''.
\newblock \href{https://dx.doi.org/10.1038/s41586-020-2587-z}{Nature {\bf 584},
  205--209}~(2020).

\bibitem{Nori_ultra_review}
Anton Frisk~Kockum, Adam Miranowicz, Simone De~Liberato, Salvatore Savasta, and
  Franco Nori.
\newblock ``\textit{Ultrastrong Coupling Between Light and Matter}''.
\newblock \href{https://dx.doi.org/10.1038/s42254-018-0006-2}{Nat. Rev. Phys.
  {\bf 1}, 19--40}~(2019).

\bibitem{Blais_cqed_review}
Alexandre Blais, Arne~L. Grimsmo, S.~M. Girvin, and Andreas Wallraff.
\newblock ``\textit{Circuit Quantum Electrodynamics}''.
\newblock \href{https://dx.doi.org/10.1103/RevModPhys.93.025005}{Rev. Mod.
  Phys. {\bf 93}, 025005}~(2021).

\bibitem{Kirchmair_nonMarkovian}
Paul Heidler, Christian M.~F. Schneider, Katja Kustura, Carlos
  Gonzalez-Ballestero, Oriol Romero-Isart, and Gerhard Kirchmair.
\newblock ``\textit{Non-Markovian Effects of Two-Level Systems in a Niobium
  Coaxial Resonator with a Single-Photon Lifetime of 10 milliseconds}''.
\newblock \href{https://dx.doi.org/10.1103/PhysRevApplied.16.034024}{Phys. Rev.
  Appl. {\bf 16}, 034024}~(2021).

\end{thebibliography}
\end{document}